\documentclass[twocolumn,letterpaper]{IEEEAerospaceCLS}

\usepackage[]{graphicx}    
\usepackage{amsfonts}
\usepackage{amsmath}
\usepackage{lipsum}
\usepackage[utf8]{inputenc}
\usepackage{booktabs}
\usepackage{makecell}
\usepackage{url}
\usepackage{multicol}
\usepackage{dblfloatfix}

\newcommand{\ignore}[1]{}

\usepackage{xcolor}

\graphicspath{{./figures/}}

\begin{document}
\title{The Parallel Coordinates Plot Revisited: Visual Extensions from Hive Plots, Heterogeneous Correlations, and an Exploration of Covid-19 Data in the United States}

\author{%
	Gary Koplik \\
	Geometric Data Analytics
	\and Ashlee Valente \\
	Geometric Data Analytics \\
}

\maketitle

\thispagestyle{plain}
\pagestyle{plain}

\begin{abstract}
This paper extends an existing visualization, the Parallel Coordinates Plot (PCP), specifically its polar coordinate representation, the \textit{Polar Parallel Coordinates Plot (P2CP)}. With the additional incorporation of techniques borrowed from Hive Plot network visualizations, we demonstrate improved capabilities to explore multidimensional data in flatland, with a particular emphasis on the unique ability to represent 3-dimensional data. To demonstrate these techniques on P2CPs, we consider toy data, the Iris dataset, and socioeconomic data for counties in the United States. We conclude with an exploration of Covid-19 data from counties in the contiguous United States.
\end{abstract}

\begin{figure}[h!]
	\centering
	\includegraphics[width=3.4in]{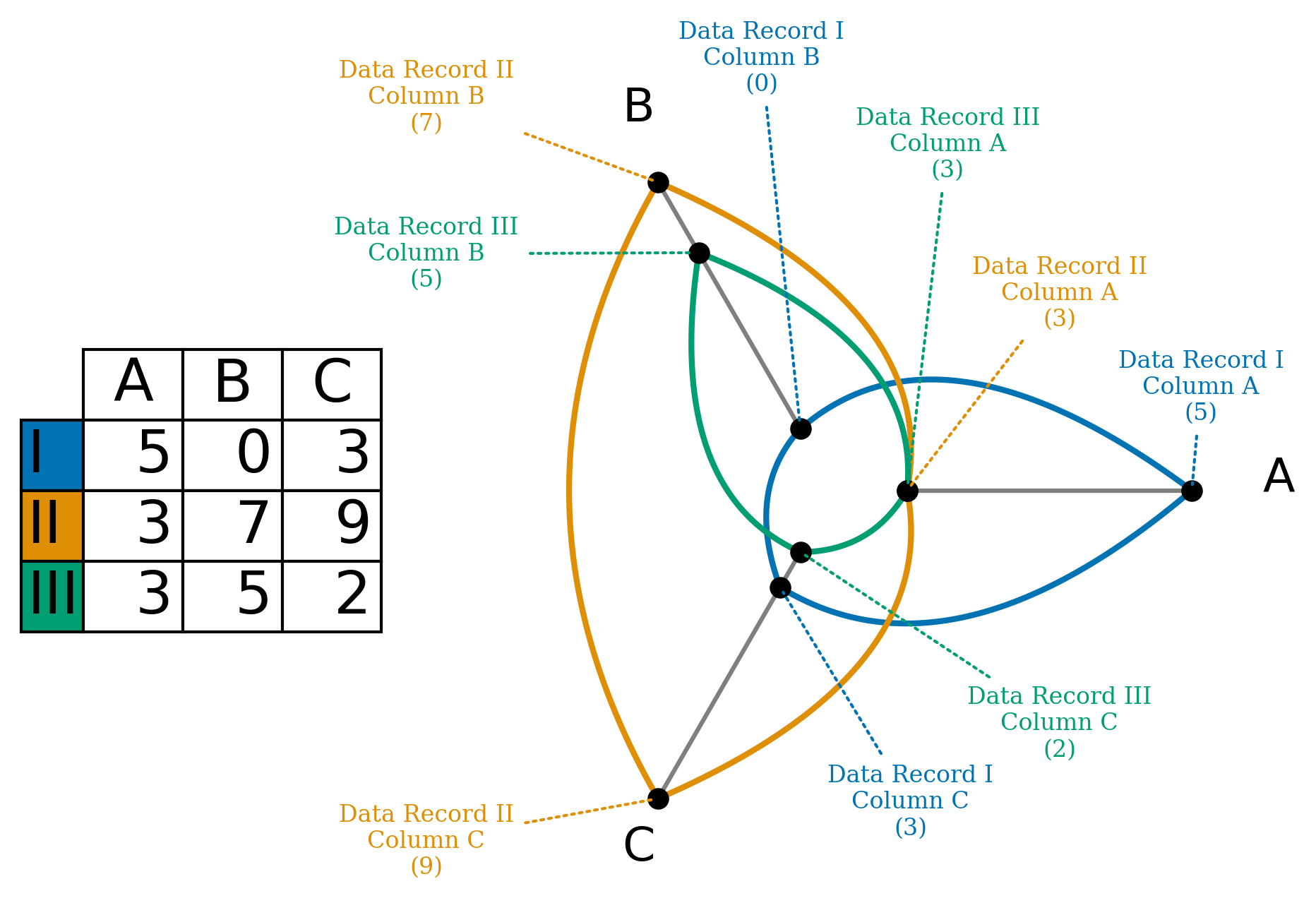}    
	\caption{A simple example of three points in three dimensions, represented as a two-dimensional Polar Parallel Coordinates Plot but styled as a Hive Plot. Each closed loop in the figure corresponds to a single row of data. Note each axis spans exactly the range of its dimension of data, and the axes therefore span heterogeneous ranges.}
	\label{fig:small_example}
\end{figure}

\section{Introduction}
\label{sec:intro}

``Reasoning about evidence should not be stuck in 2 dimensions, for the world we seek to understand is profoundly multivariate'' --- Edward Tufte \cite{tufte2006beautiful}.

Multidimensional data visualization is routinely stuck between two conflicting truths---humans are very good at finding patterns in data that can be visualized in two dimensions, but plotting high-dimensional data in 2d requires representational tricks that can come with an interpretability trade-off.

Plenty of clever techniques have been applied to comprehensibly reduce multidimensional data into flatland. Perhaps the simplest technique used is to look at all of the pairs of bivariate Scatterplots in a Scatterplot Matrix. Another commonly-used technique is to plot a dataset as a standard bivariate Scatterplot using two of the available dimensions, but modify the appearance of each plotted point according to additional dimensions of data. For example, the color and size of each point can represent extra dimensions, with techniques like Chernoff faces \cite{chernoff1973use} extending this concept to higher dimensions. Others prefer to use embedding techniques such as Principal Components Analysis to explicitly collapse the information content of multidimensional data into flatland. Finally, assigning dimensions to \textit{axes} in flatland, a technique most notably used with the Parallel Coordinates Plot (PCP), allows one to represent a multidimensional point as a \textit{polyline} while preserving univariate relationships as well as a subset of bivariate relationships.

Just as PCPs visualize relationships between orthogonal dimensions of multivariate data, a \textit{Hive Plot (HP)} visualizes relationships between orthogonal sets of nodes in network data. In this paper, we delve into the overlap between these two visualization techniques, applying concepts from HPs to PCPs in order to improve the analytical capabilities of PCPs. We demonstrate the resulting visualization, the \textit{Polar Parallel Coordinates Plot (P2CP)}, on toy data as well as the Iris dataset, county-level socioeconomic data for the United States, and county-level Covid-19 data from the contiguous United States.

\subsection{Outline}

We first discuss the history of Parallel Coordinates Plots and Radar Charts in Section \ref{sec:pcp}. We then apply techniques from the Hive Plot literature on several example datasets in Section \ref{sec:hive_plots}. Next, we visualize correlations in Polar Parallel Coordinates Plots in Section \ref{sec:corr}. As an application of Hive Plot techniques and exploration of heterogeneous correlations, we look at county-level Covid-19 data for the contiguous United States in Section \ref{sec:covid}, and we conclude in Section \ref{sec:conclusion}.

\section{Parallel Coordinates Plots and Polar Doppelgangers}
\label{sec:pcp}

The Parallel Coordinates Plot (PCP) extends as far back as the 1880s. Philbert Maurice d'Ocagne in 1885 is frequently credited with laying out the means of coordinate transformation needed for PCPs \cite{d1885coordonnees}, but Henry Gannett preceded d'Ocagne with the first plots a few years earlier \cite{hewes1883scribner}. PCPs were popularized in 1980s by Alfred Inselberg \cite{inselberg1985plane}. Heinrich and Weiskopf \cite{heinrich2013state} nicely summarize many of the recent innovations in PCPs.

The polar coordinate equivalent of PCPs, the \textit{Polar Parallel Coordinates Plot} (P2CP), which radiates axes out from the origin, is essentially an applied use of the Radar Chart (RC), also known as the Spider Plot or Star Plot. RCs actually precede PCPs by several years, with Georg von Mayr publishing the first RC in 1877 \cite{von1877gesetzmassigkeit}.

Although some researchers worry that PCPs can be hard to interpret, there is evidence that even the unfamiliar can in fact understand PCPs \cite{siirtola2009visual}. Furthermore, Lazenberger et al. \cite{lanzenberger2005exploring} found evidence suggesting that although PCPs were more interpretable at first glance, P2CPs were better for hypothesis generation as well as the visualization of important insights.

Unfortunately, although RCs are a well-established visualization technique, there is minimal exploration in the literature of their use as P2CPs, with only some discussion in the context of \textit{Circular Parallel Coordinates Plots} \cite{hoffman2000table} and \textit{Stardinates} \cite{lanzenberger2003interactive}. Instead, RCs, both in the literature and in practice, focus on \textit{shape comparisons} between only a few multidimensional points rather than considering a far larger number of data points as one usually shows with a PCP. This limited use of RCs has many flaws. For example, many RC tools fill in the area inside the completed loop representation of a multidimensional data point, which leads to a visually disingenuous quadratic increase in area with respect to a linear increase along one dimension. Furthermore, there are far more interpretable visualizations than RCs for comparing so few multidimensional points. For a more thorough discussion critiquing RCs, see \cite{radarchartcritique}.

By borrowing a few concepts from Hive Plots---a technique from network visualization---this paper extends the visualization capabilities of P2CPs to larger datasets. In particular, we focus on curved edges, the unique visualization power of a 3-axis P2CP, and small multiples.

\section{Visualization Techniques from Hive Plots}
\label{sec:hive_plots}

HPs \cite{krzywinski2012hive} serve as an interpretable means of network visualization. Due to the abstract nature of most networks, there usually is not a ``correct'' way to place nodes in Euclidean space. Whereas many network visualizations rely on an algorithmic placement of nodes that can be difficult to interpret, HPs allow one to more explicitly position nodes. First, one chooses a partitioning of the nodes, which consequently dictates a set of distinct \textit{axes} on which to place each set of nodes. The user then selects a \textit{scalar ordering variable} for each axis. A node is thus precisely and interpretably placed on its specified axis in two-space based on its corresponding scalar value. After drawing edges between nodes in their final placement, the resulting structure offers a strong visual interpretability, suggesting anecdotal patterns between nodes with respect to the sorting variables chosen by the user.

To connect P2CPs to HP visualization techniques, we must first clarify how we will think of standard multidimensional data as ``network data'' of a sort. Rather than thinking of choosing our axes to be some partition of the nodes in a network, we will let our axes be dimensions of the dataset. We will thus think of a ``node'' on an axis as being a specific dimension value in a record of data. Assuming we do not have missing values in our data, one multidimensional data point will thus consist of one closed loop in a HP. A simple example of three points in three dimensions is shown in Figure \ref{fig:small_example}.

We borrow several concepts from HP visualizations to extend the capabilities of P2CPs with many data points.

\subsection{Curvature of Edges}

\begin{figure}[h!]
	\includegraphics[width=3.4in]{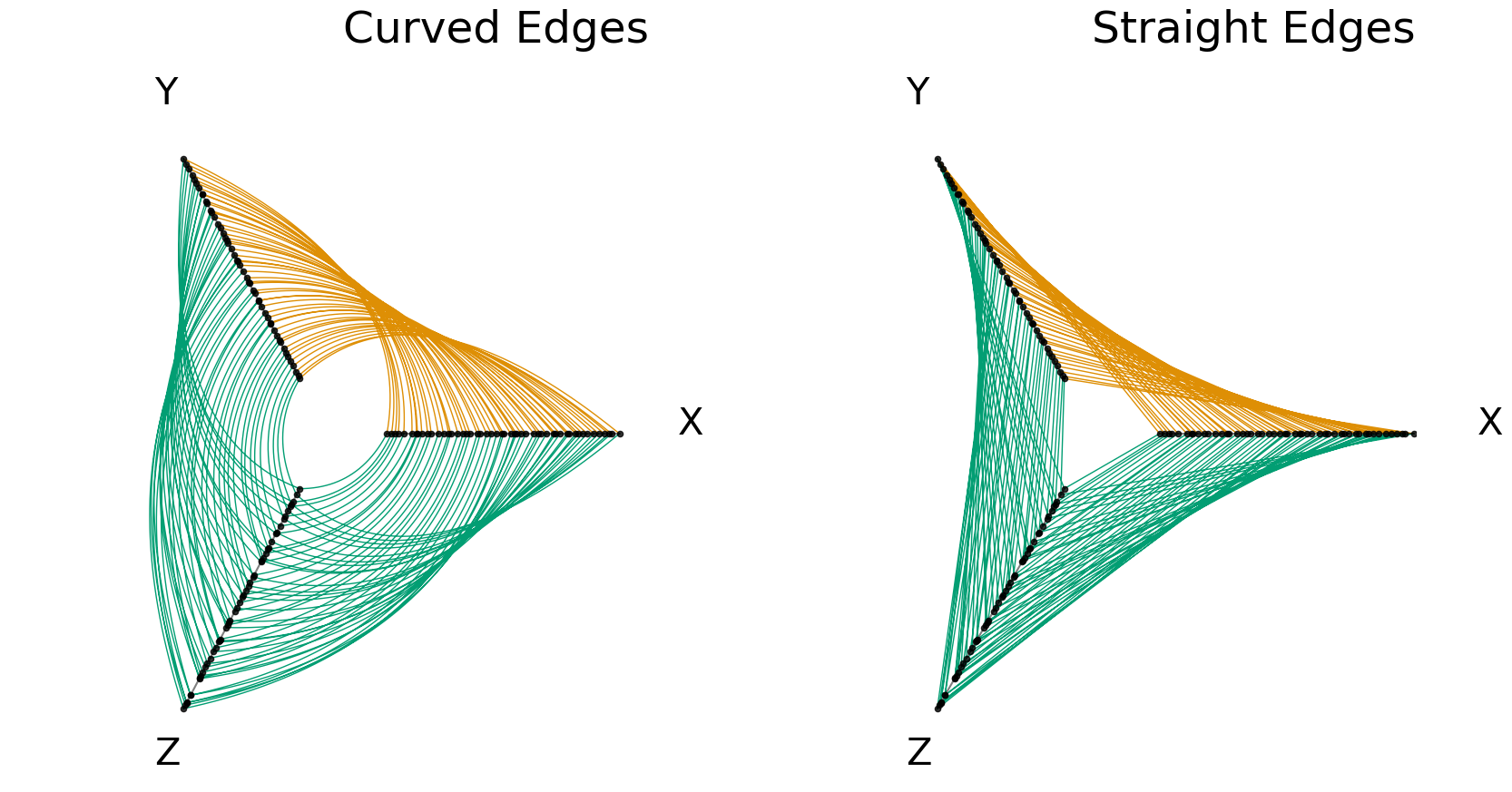}    
	\caption{An example of two types of bivariate relationships represented in 3-dimensional Polar Parallel Coordinates Plots with either curved (left) or straight (right) edges. The $X$ to $Y$ (orange) relationship represents a strong negative correlation while the $X$ to $Z$ and $Y$ to $Z$ (green) relationships represent a heterogeneous correlation. Visualization of correlations with Polar Parallel Coordinates Plots is discussed in detail in Section \ref{sec:corr}, but for now we simply note that with straight edges, the lines are relatively compressed and hard to distinguish. Furthermore, we observe a misleading, visually-implied concavity with straight edges that does not appear with curved edges.}
	\label{fig:straight_vs_curved_edges}
\end{figure}

RCs and P2CPs almost always follow the PCP convention of straight line connections between axes,\footnote{There are some exceptions of PCPs contemplating curved edges, for example \cite{heinrich2011evaluation}.} thus under-utilizing the available space in polar coordinates. HPs on the other hand draw B\'ezier curves that arc through the space.\footnote{One notable Radar Chart tool with curved edges is available through Google Docs, which draws line arcs quite similarly to the authors' preferred B\'ezier curve structure. Examples can be found at https://support.google.com/docs/answer/9146868} The curved edges not only make better use of the space in a plot in polar coordinates, but also avoid visual artifacts of concavity that can appear in figures with a more dense number of edges, exemplified in Figure \ref{fig:straight_vs_curved_edges}.

\subsection{3 Dimensions Preserved in Flatland}

\begin{figure}[h!]
	\centering
	\includegraphics[width=2.in]{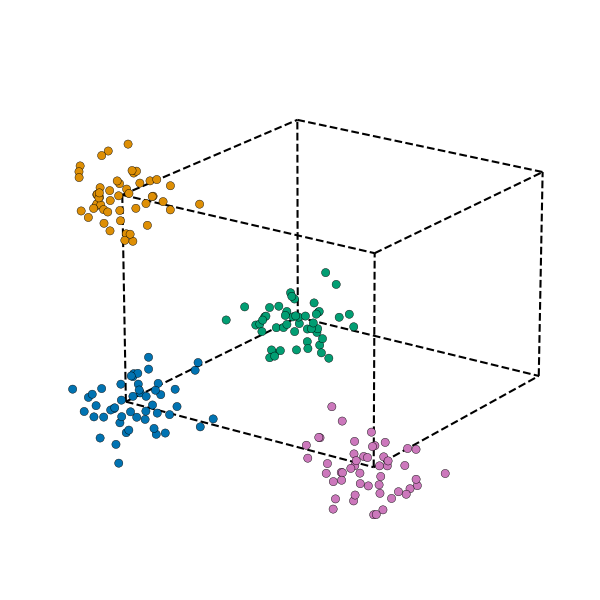}    
	\caption{Four Gaussian blobs centered on four corners of a cube.}
	\label{fig:cube_3d}
\end{figure}

Static visualizations of 3-dimensional data can be highly misleading. As an example, consider the following toy dataset---four Gaussian blobs centered at four different corners of a cube (Figure \ref{fig:cube_3d}).

\begin{figure}[h!]
	\centering
	\includegraphics[width=3.4in]{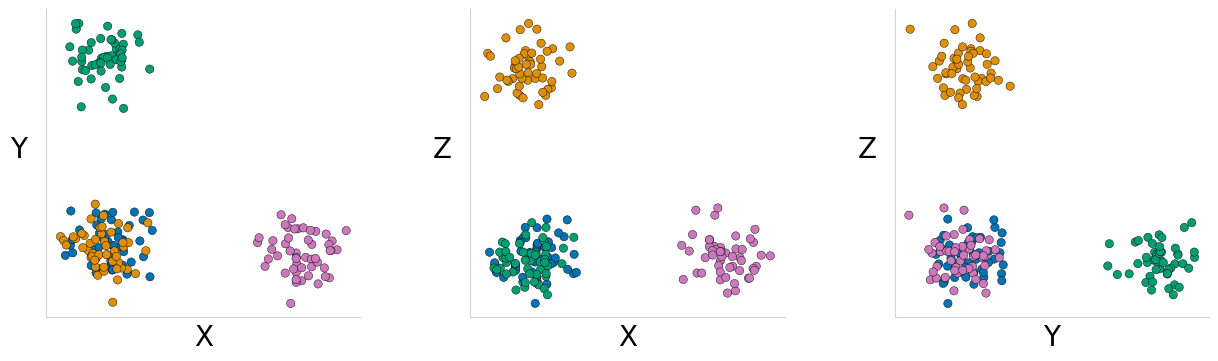}    
	\caption{The same four Gaussian blobs from Figure \ref{fig:cube_3d}, looking directly at the $XY$, $XZ$, and $YZ$ planes. One can only distinguish the four clusters by looking at more than one of these figures. The separability is also far less evident (for example, the blue cluster is never fully isolated in any figure).}
	\label{fig:cube_scatterplots}
\end{figure}

With the strategic choice of angle and azimuth in Figure \ref{fig:cube_3d}, the four clusters are clearly distinguished, but if we look at the wrong angle, the separability becomes far less visually apparent (Figure \ref{fig:cube_scatterplots}).

\begin{figure}[h!]
	\centering
	\includegraphics[width=2.in]{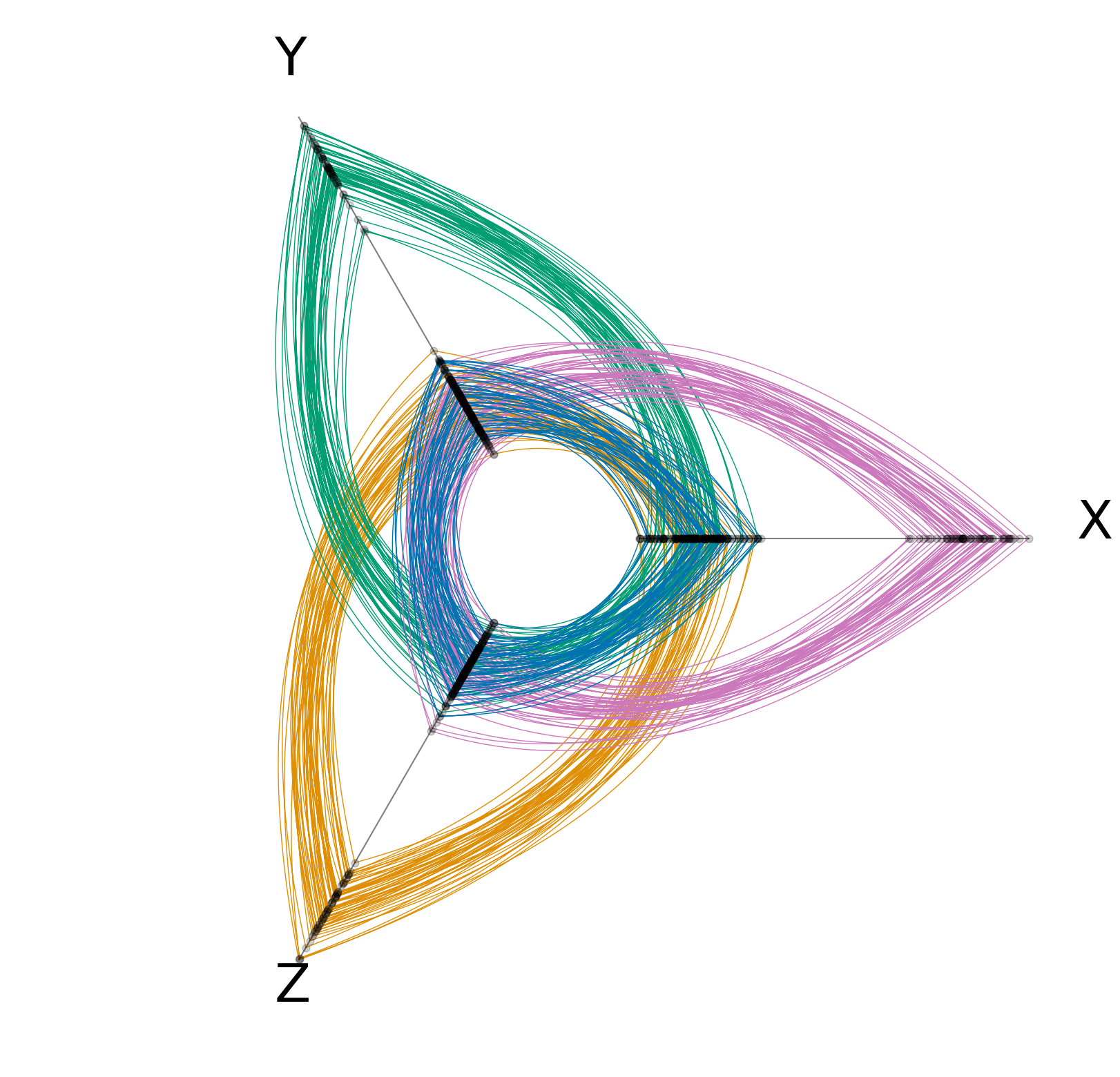}    
	\caption{The same four Gaussian blobs from Figure \ref{fig:cube_3d}, represented as a Polar Parallel Coordinates Plot but styled as a Hive Plot. One can now easily distinguish the four clusters by looking at the shape of the loops formed over the three dimensions. Note that in addition to showing this trivariate relationship, we also show \textit{every} bivariate and univariate relationship without any repeat axes.}
	\label{fig:cube_hp}
\end{figure}

If we instead place our $X$, $Y$, and $Z$ axes in flatland with a 3-axis P2CP (Figure \ref{fig:cube_hp}), we can see the unambiguous trivariate separability, as exhibited by the appearance of four loops of distinct shape and color . Though 3-axis P2CPs can of course nicely demonstrate trivariate relationships, the use of three axes has a particular motivation for HPs.

HPs typically have three axes because \textit{each axis is adjacent to every other axis}, and thus one would never need to draw an edge between nodes that crosses over an axis. Given one might have any arbitrary partitioning of nodes into groups, this enables cleanly visualizing any of the possible bivariate relationships in the resulting figure.

When translating this property to P2CPs, a 3-axis P2CP results in \textit{every} bivariate relationship being visually represented in a single two-dimensional figure.\footnote{P2CPs also preserve univariate relationships on each axis.} Therefore, despite projecting onto flatland, \textit{3-dimensional P2CPs do not collapse any variable interactions in the resulting figure}.

As the number of dimensions increases beyond 3, PCPs and P2CPs are still perfectly valid visualization schemes, but it should be noted that new issues arise that can affect interpretability. In particular, both the ordering of axes \cite{zhang2012network} and strategic placement of repeat axes \cite{hurley2010pairwise} are research questions in their own right.

\begin{figure}[h!]
	\centering
	\includegraphics[width=3.4in]{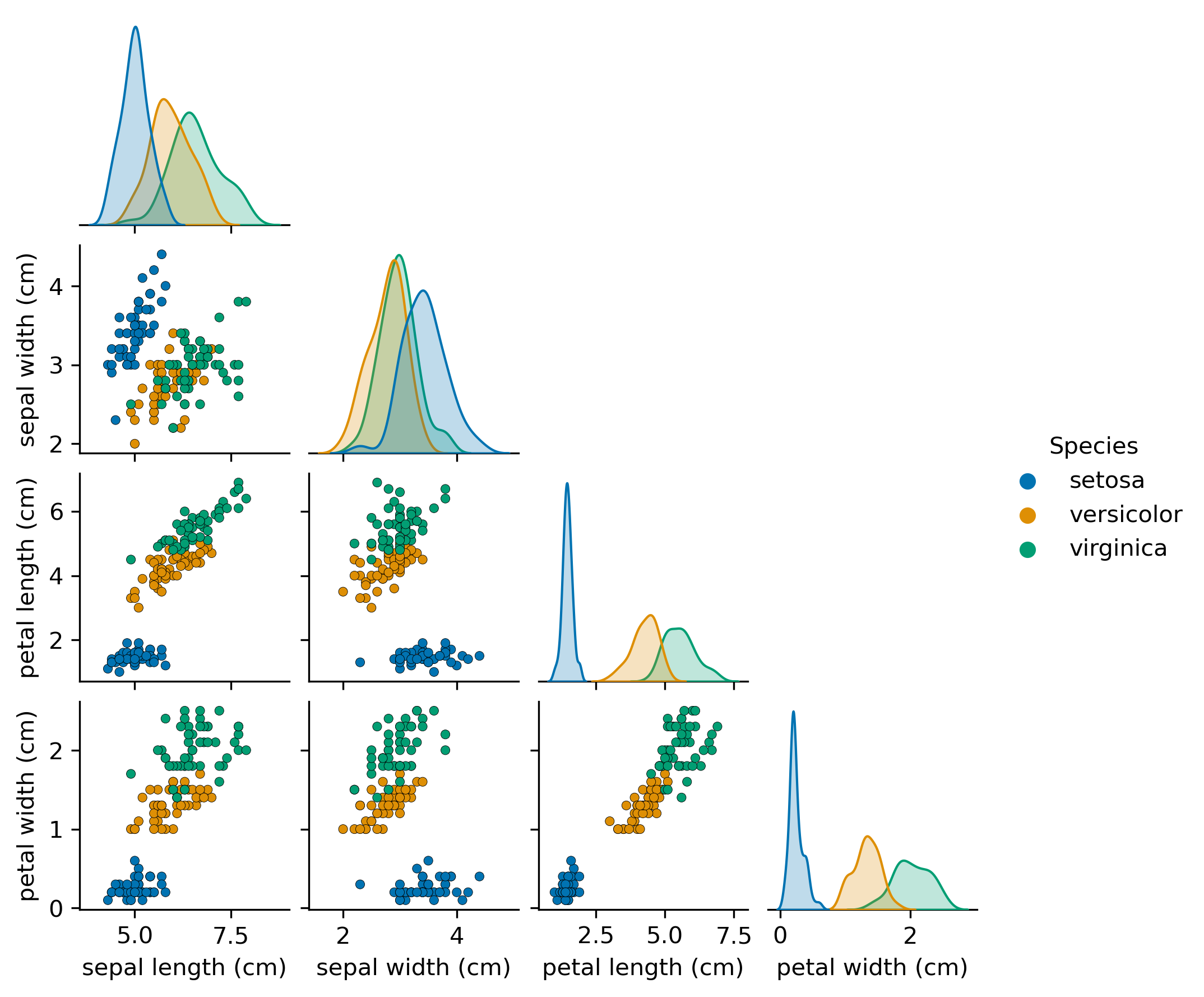}    
	\caption{A Scatterplot Matrix for the Iris dataset. The vertical separability of colors visible on the bottom two rows indicates that petal length and petal width are good predictors of flower type. Furthermore, the bottom-right Scatterplot indicates that these two variables are highly correlated. Though this implies we will have one particularly strong first Principal Component, we cannot easily infer if the Iris types can be further separated by additional Principal Components.}
	\label{fig:iris_scatter}
\end{figure}

If forced to look at a single, static visualization, a 3-axis P2CP is in fact \textit{more informative} than a 3-dimensional Scatterplot. Consider the Iris Dataset \cite{anderson1936species} \cite{fisher1936use}, which consists of three types of flowers represented by four dimensions of data. The labels separate quite well on both petal length and petal width, as demonstrated by a standard Scatterplot Matrix shown in Figure \ref{fig:iris_scatter}.\footnote{As a digression of examples using P2CPs, the Scatterplot Matrix in Figure \ref{fig:iris_scatter} could instead be represented as four instances of 3-axis P2CPs showing the four possible trivariate relationships. The resulting figure can be found in the Appendix (Figure \ref{fig:iris_p2cp_panel}).}

\begin{figure*}[h!]
	\centering
	\includegraphics[width=3.in]{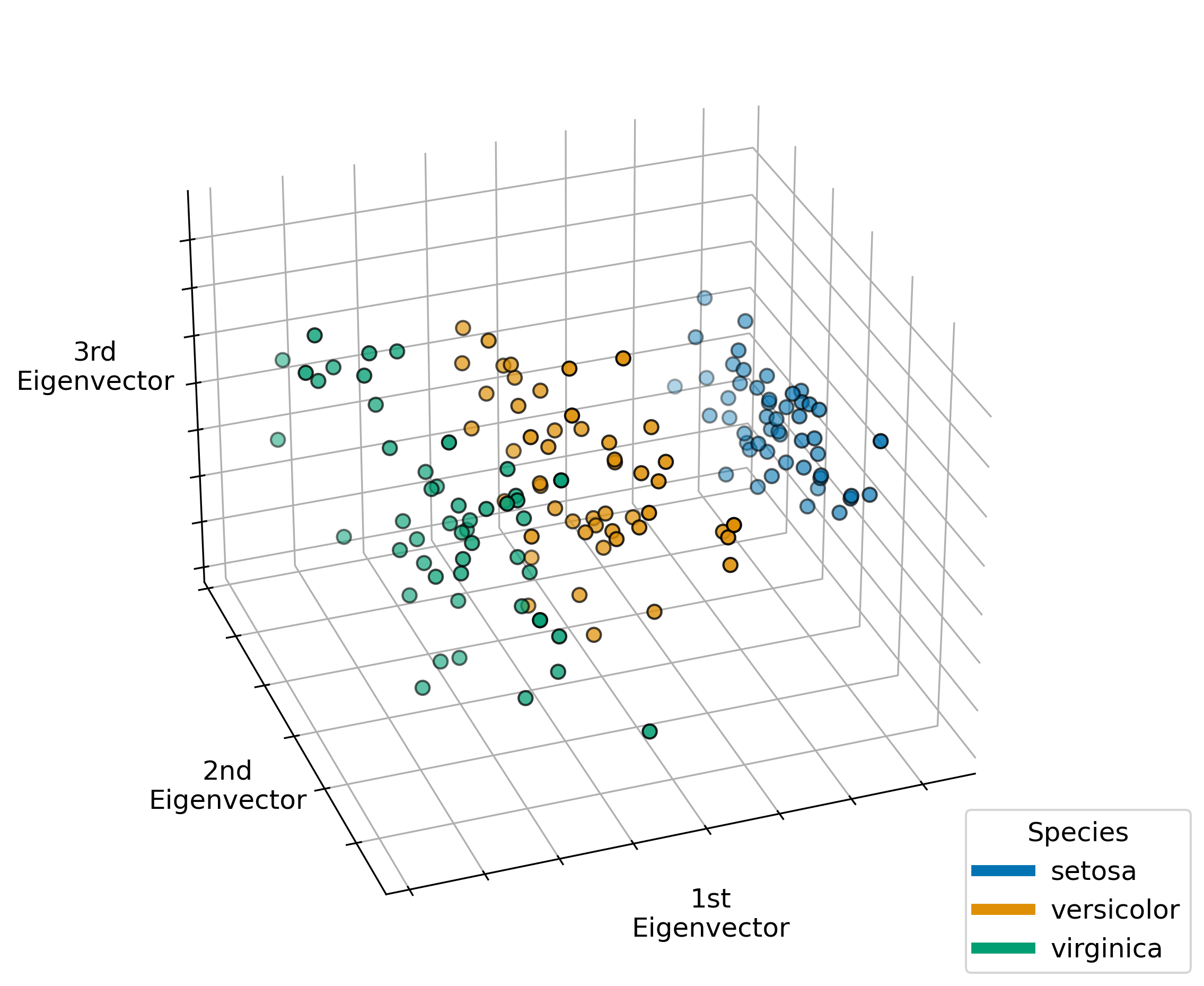} \qquad \includegraphics[width=3.in]{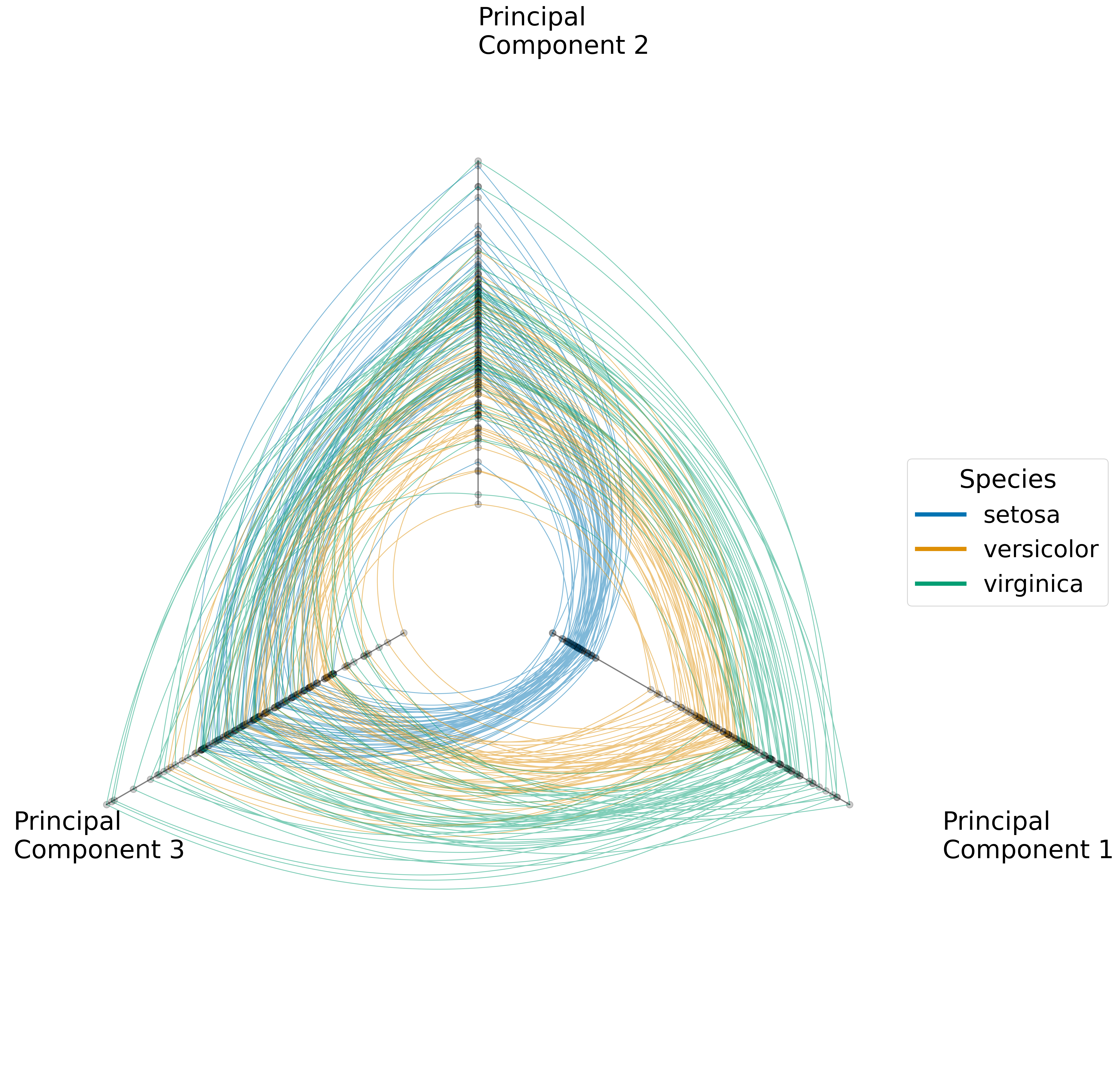} 
	\caption{\textit{Left}: 3d Scatterplot of the first three Principal Components for the Iris dataset. The first component clearly separates the data quite well, but we cannot say anything about the second or third components, let alone any combinations of components, from this 3d visualization alone. \textit{Right}: Polar Parallel Coordinates Plot representation of the first three Principal Components of the Iris dataset (the same data as shown in the left plot), styled as a Hive Plot. In a single figure in flatland, we can see all three univariate relationships, the three possible bivariate relationships, and the sole trivariate relationship. From this figure, we can conclude that the only separability for labels in the Iris dataset is along the first Principal Component.}
	\label{fig:iris_pca}
\end{figure*}

Suppose we simply want to explore the separability of the different labels via Principal Components Analysis (PCA). Looking at the first three Principal Components (PCs) in three dimensions (the left plot in Figure \ref{fig:iris_pca}), we can only draw one conclusion---the first PC performs quite a bit of separation on its own---but we are unable to conclude anything further. In fact, there is no single 3d Scatterplot of the PCA dataset that can unambiguously show us the collective separability possible using these three PCs.

A 3-axis P2CP visualization of the same dataset (the right plot in Figure \ref{fig:iris_pca}), on the other hand, not only demonstrates the standalone, univariate separability resulting from PC1, but also shows us that the second and third PCs have no univariate separability. Furthermore, this visualization illustrates that no bivariate relationship or even the one trivariate relationship can improve on the separability achieved by PC1 alone.


\begin{figure*}[hb!]
	\centering
	\includegraphics[width=7in]{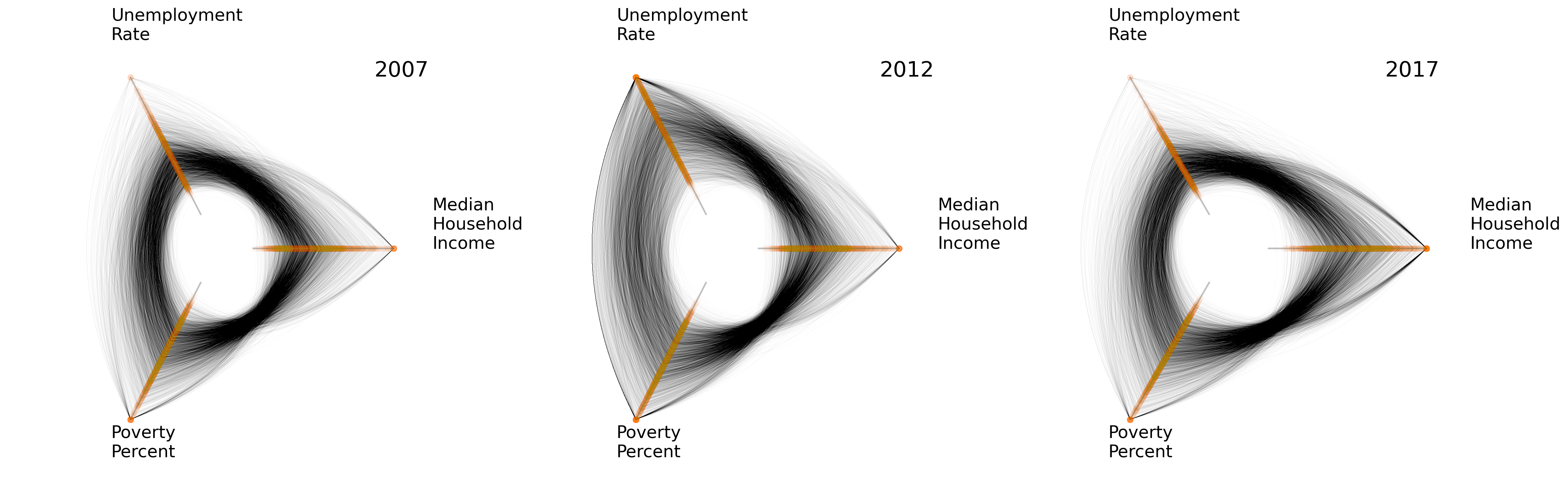}    
	\caption{Polar Parallel Coordinates Plots of unemployment rate, poverty percentage, and median household income for United States counties in 2007, 2012, and 2017. Univariate density is represented with the orange dots on each axis in the figure. The bivariate correlation between unemployment and poverty appears consistently positive over the represented years, notably with unemployment and poverty increasing in tandem during the economic fallout from the Great Recession. Similarly, the negative correlation between poverty and median household income appears robust over the represented years, though median household income values increase over time (note, these data are not inflation-adjusted). The last bivariate correlation between unemployment and median household income shows an intriguing visual heterogeneity over time. Despite the Pearson correlation being relative consistent over the three represented years ($-0.36$, $-0.40$, and $-0.41$ for 2007, 2012, and 2017 respectively), there appear to be changing trends in behaviors along the upper extremes between these two variables. In 2007, both high unemployment and income appear to be \textit{symmetrically} negatively correlated with each other, but in 2012, the negative relationship appears to be stronger for high unemployment than for high income. This relationship then appears to flip in 2017, with high income having a visually stronger negative relationship than for high unemployment. Visualizing correlations with Polar Parallel Coordinates Plots is discussed in more detail in Section \ref{sec:corr}. Note that the minimum and maximum values on the axes of each plot have been constrained to normalize outliers to a more tightly-bound range.}
	\label{fig:unemployment}
\end{figure*}

\begin{figure*}[bh!]
	\centering
	\includegraphics[width=3.4in]{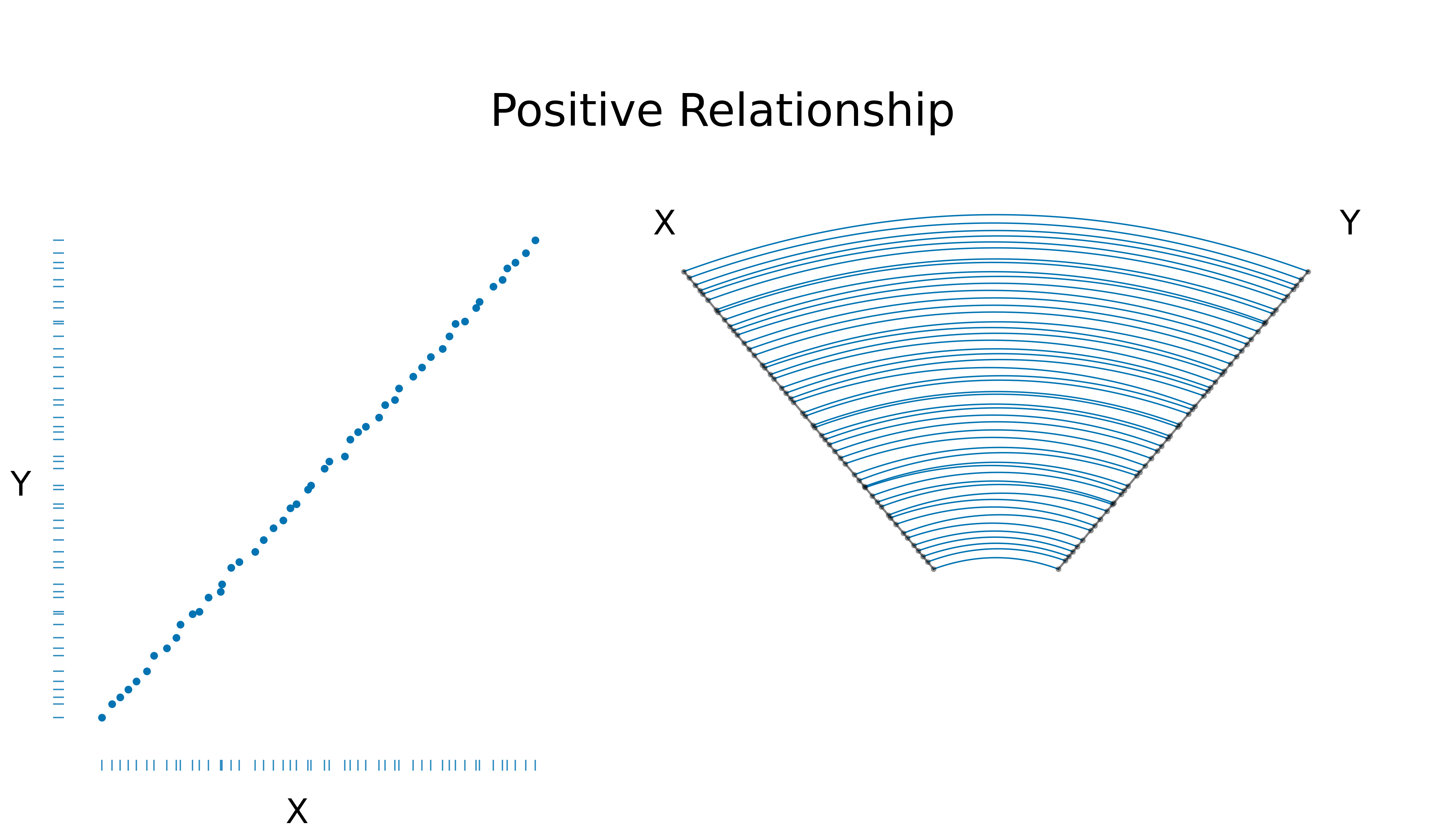} \includegraphics[width=3.4in]{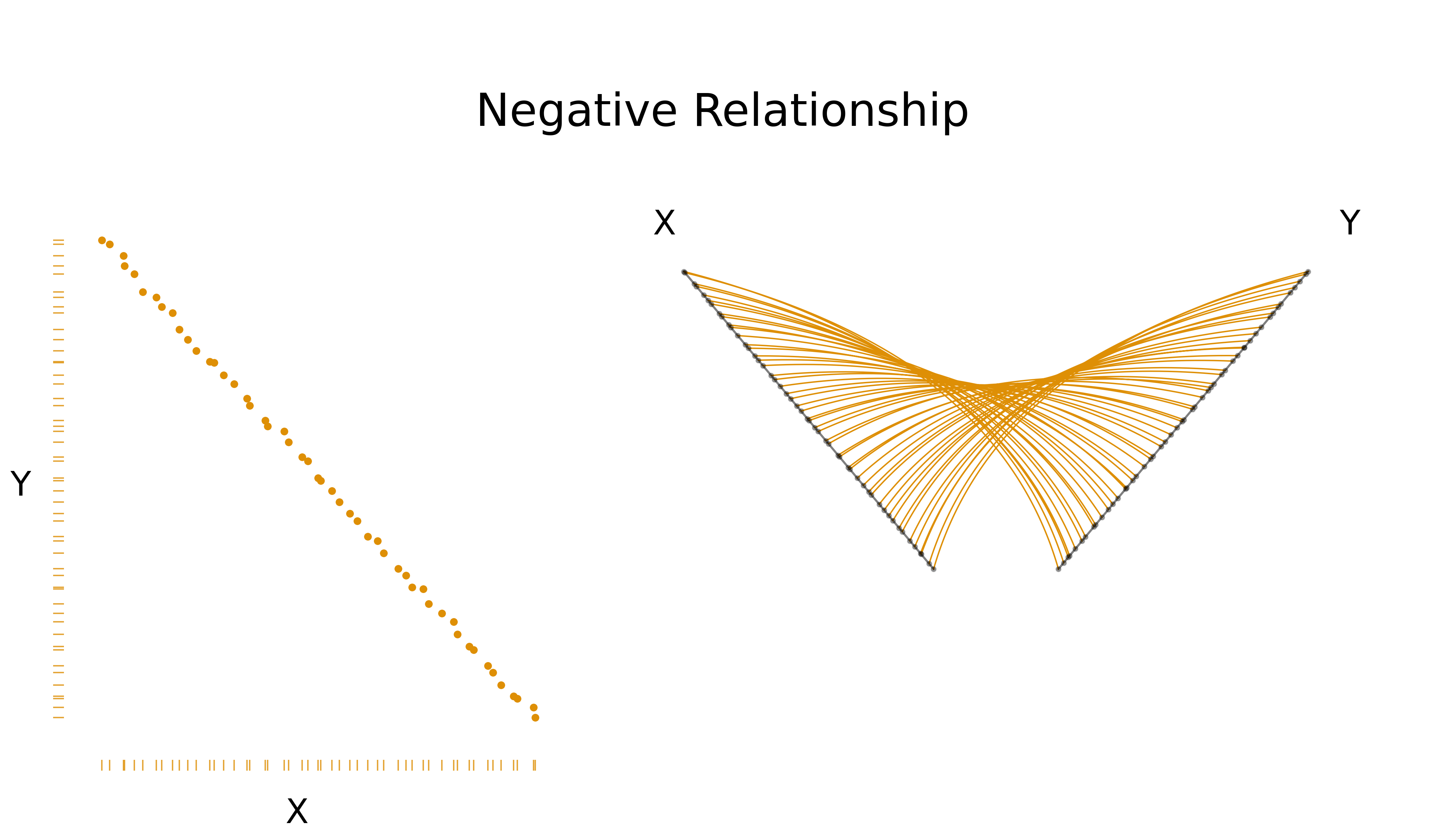} 
	\includegraphics[width=3.4in]{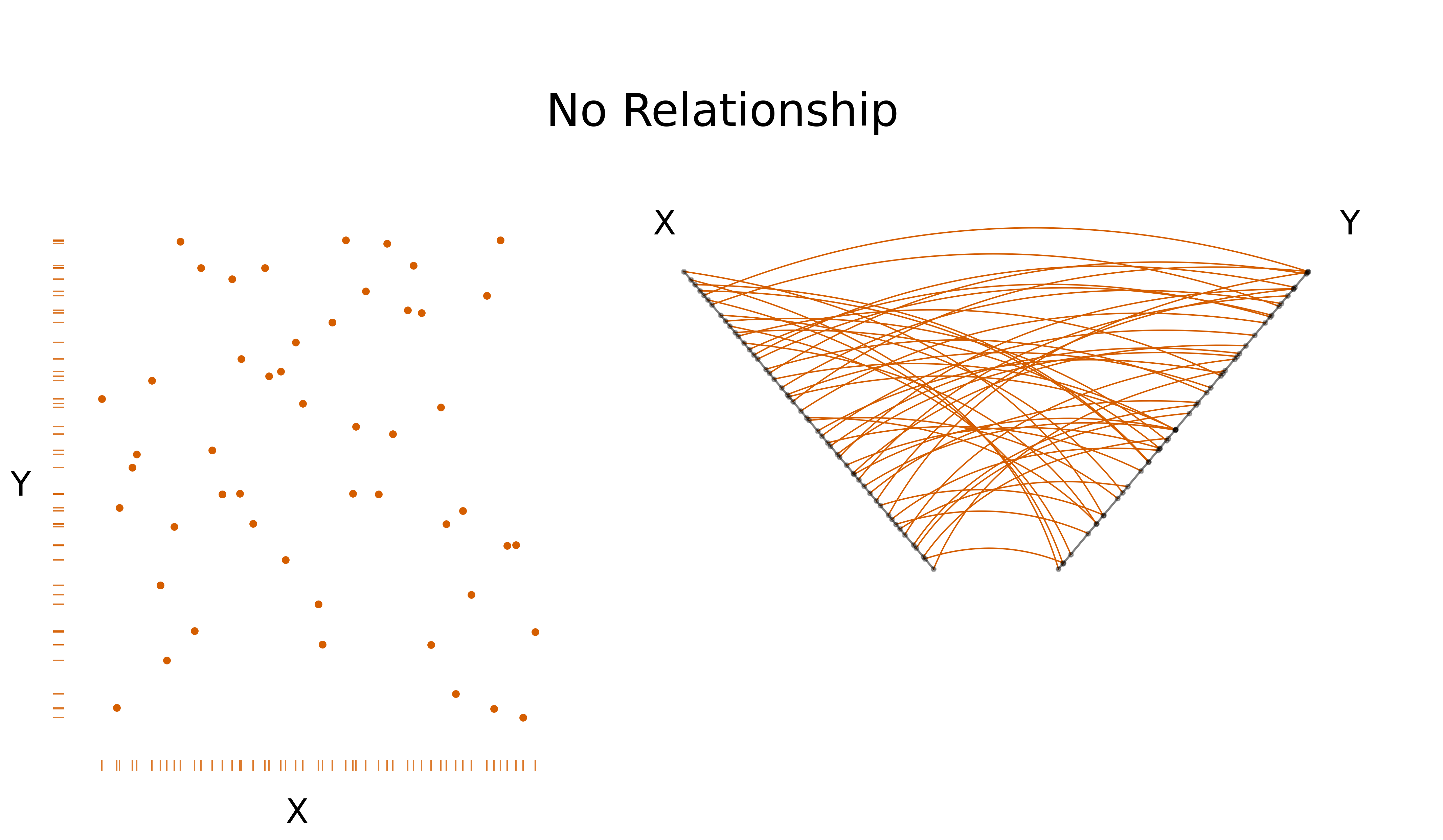} \includegraphics[width=3.4in]{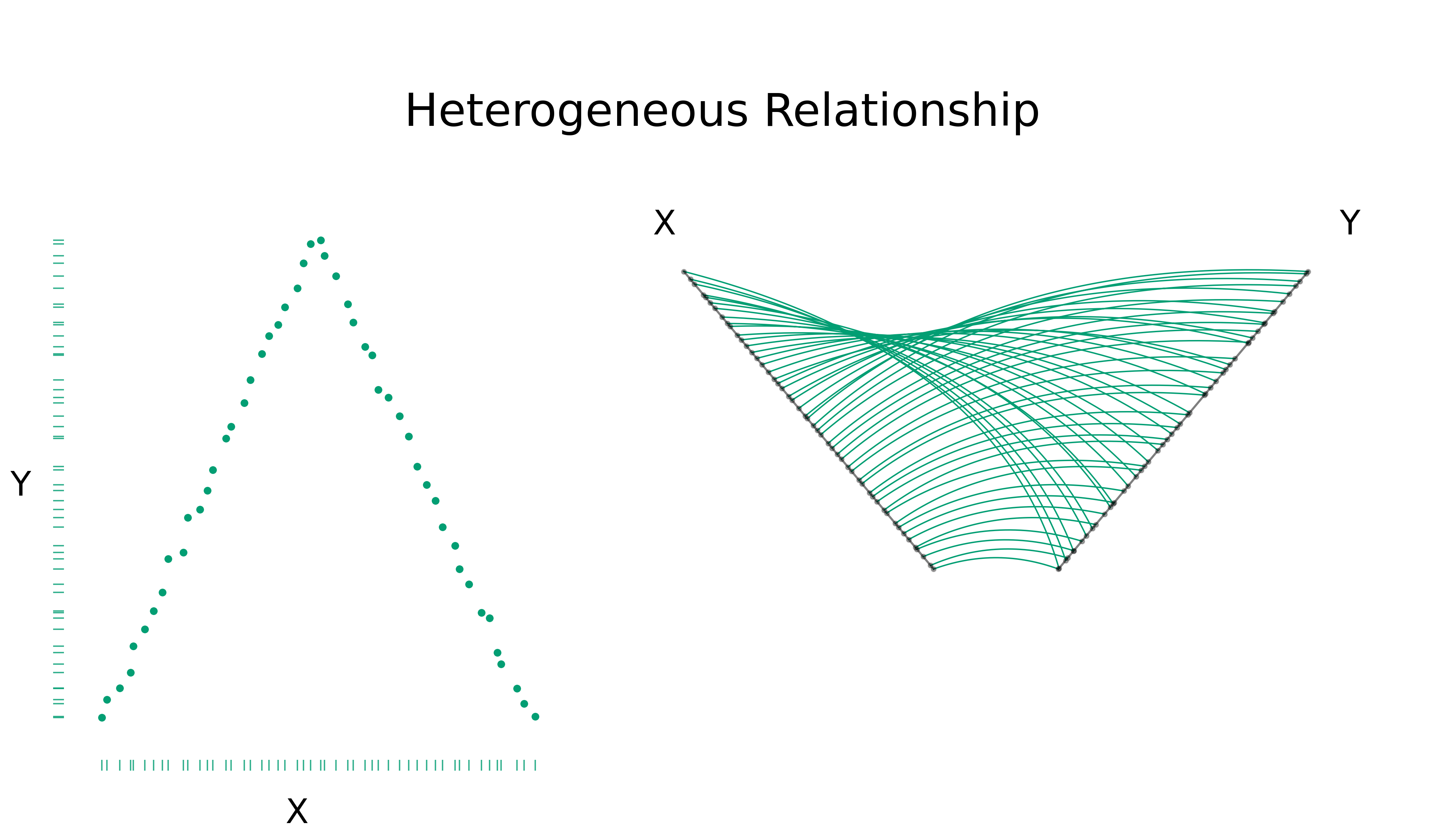}   
	\caption{Four simple examples of different correlations. At the left of each figure, a Dot-Dash Plot shows the data (e.g. a bivariate Scatterplot showing univariate density on each axis). On the right of each figure, the comparable Polar Parallel Coordinates Plot representation of the same data. Note the negative and heterogeneous relationships in this figure were also used in Figures \ref{fig:straight_vs_curved_edges} and \ref{fig:trivariate_corr}.}
	\label{fig:bivariate_corr_examples}
\end{figure*}

\subsection{Small Multiples with Hive Panels}

The compact nature of HPs allows one to look at \textit{small multiples} of plots in succession with each other, referred to in the HP literature as a \textit{Hive Panel} \cite{krzywinski2012hive}. Although small multiples have frequently been used with RCs, they have not been used in the PCP literature. One use case for small multiples with parallel coordinates would be to visualize multiple combinations of dimensions from a high-dimensional dataset---for example, instead of making a single, long PCP with $n$ axes, one could make $n/3$ P2CPs, each with three axes.

A second natural use case arises with multidimensional time series data, for which one might be interested in visualizing the changing multivariate relationship over time. Current methodologies in the literature for showing changes over time with PCPs include adding axes for discrete times \cite{dietzsch2009spray}, using color and alpha channels to continuously correspond with time \cite{johansson2007depth}, and animation tools over the same set of axes \cite{barlow2004animator} \cite{theron2006visual}. With P2CPs, we can instead partition the data over discretized windows of time and look at the resulting P2CPs side-by-side. As an example, in Figure \ref{fig:unemployment}, we generated a series of P2CPs with three socioeconomic variables using data from the Bureau of Labor Statistics \cite{countyUnemploymentData} and the United States Census Bureau \cite{countyPovertyData}---unemployment rate, poverty percentage, and median household income---for United States counties in 2007, 2012, and 2017. This figure shows some consistent relationships over time, most notably a stable negative correlation between poverty and income, but also demonstrates an intriguing heterogeneity over time between unemployment and income.

\section{Visualizing Multivariate, Heterogeneous Correlations with Polar Parallel Coordinates Plots}
\label{sec:corr}

A large part of Exploratory Data Analysis involves looking for \textit{patterns}, with a particularly important pattern between two variables being their \textit{correlation}. In this section, we look at basic bivariate correlations in P2CPs and then turn our attention to \textit{multivariate} correlations. We demonstrate with a simple example how in a circumstance of \textit{heterogeneous correlations}, the multivariate visualization scheme of P2CPs lends itself well to finding \textit{localized, multivariate patterns} in the data, a strategy we will make use of in practice in Section \ref{sec:covid}.

Before discussing the visualization of correlations with P2CPs further, it's important to note that when visually exploring bivariate correlations, there are known interpretability trade-offs when using PCPs as opposed to Scatterplots, discussed further in \cite{li2010judging}, \cite{johansson2008perceiving}, and \cite{kuang2012tracing}. These trade-offs, however, extend only to the simpler task of identifying the magnitude of positive and negative correlations as noise increases. For the majority of this section, we will focus on \textit{heterogeneous} correlations and the particular capability with P2CPs to discern localized, multivariate patterns, a task that we consider outside the scope of the above-cited papers.

To demonstrate basic correlation patterns with P2CPs, we first consider four simple, low-noise examples in Figure \ref{fig:bivariate_corr_examples}. Dot-Dash Plots \cite{tufte2001visual} were used for the bivariate Scatterplots to create a more comparable visualization to the P2CP, as P2CPs also display univariate density on their axes. For P2CPs, positive relationships result in concentric arcs, negative relationships result in a ``butterfly'' pattern, and no relationship conveniently looks like no relationship.

\begin{figure}[h]
	\centering
	\includegraphics[width=3.in]{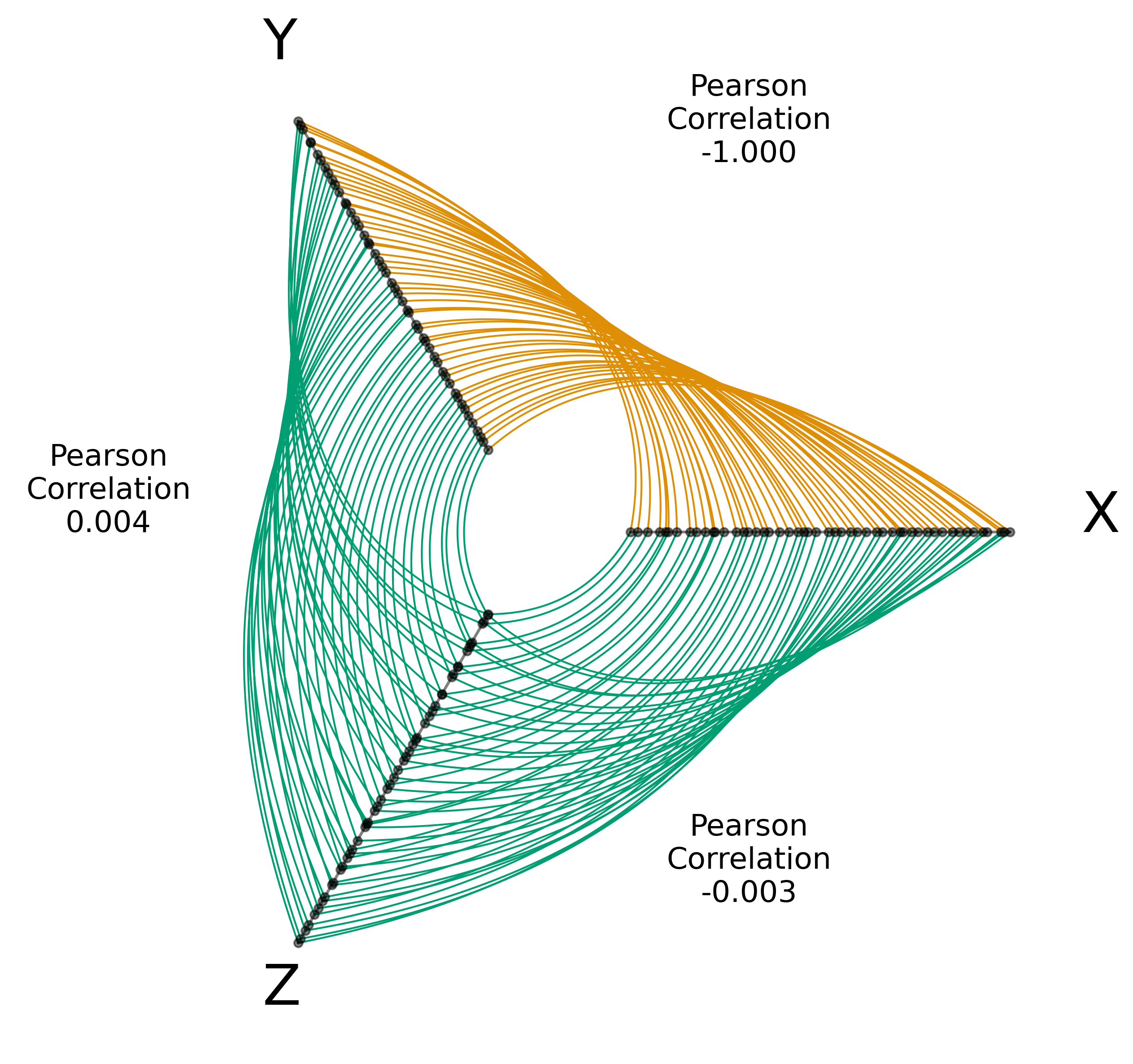}    
	\caption{An example trivariate Polar Parallel Coordinates Plot, with one negative relationship ($X$ to $Y$) and two heterogeneous relationships ($X$ to $Z$ and $Y$ to $Z$) borrowed from Figure \ref{fig:bivariate_corr_examples}. By design, the heterogeneous relationships here contain subsets with either positive or negative correlations, resulting in overall correlations of approximately 0. Despite no overall correlation for the heterogeneous relationships, this visualization allows us to visualize local, multivariate patterns in this toy dataset. For example, a high value for any one variable relates to a lower value for the other two variables.}
	\label{fig:trivariate_corr}
\end{figure}

Heterogeneous correlations can of course vary drastically from our simple example in Figure \ref{fig:bivariate_corr_examples}, but regardless of the particular heterogeneous structure, one can likely discern \textit{local patterns} in that bivariate relationship with a standard Scatterplot as well as with a P2CP representation. With P2CPs though, one can additionally explore those local patterns \textit{in a multivariate context}. As a simple toy example, consider a 3-variable toy dataset in Figure \ref{fig:trivariate_corr} composed of two heterogeneous relationships and one negative relationship, borrowing correlation structures from Figure \ref{fig:bivariate_corr_examples}. With this visualization, we can quickly discern multivariate patterns; for example, a high value for any one variable relates to a lower value for the other two variables.

\begin{figure*}[bh!]
	\centering
	\includegraphics[width=2.3in]{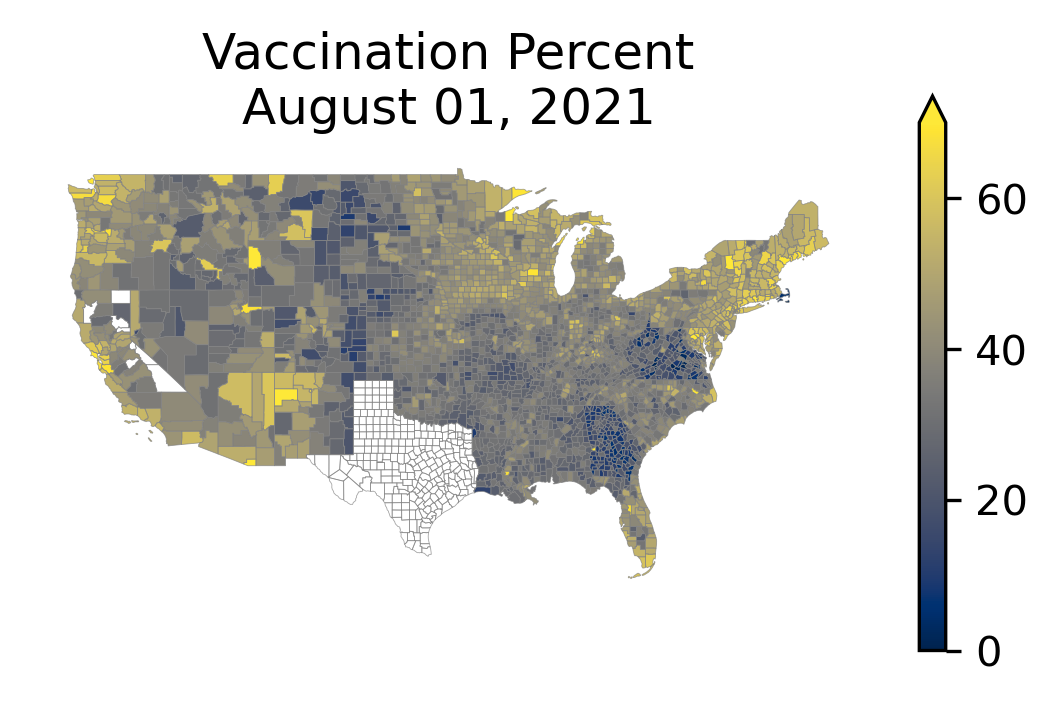} \includegraphics[width=2.3in]{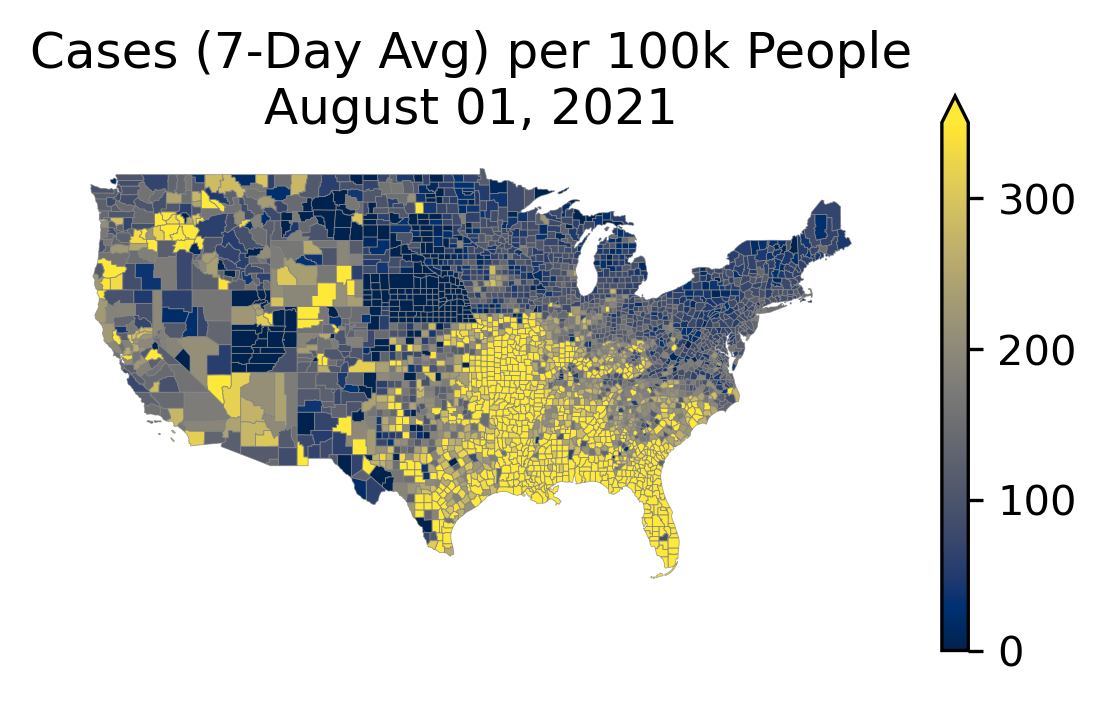} \includegraphics[width=2.3in]{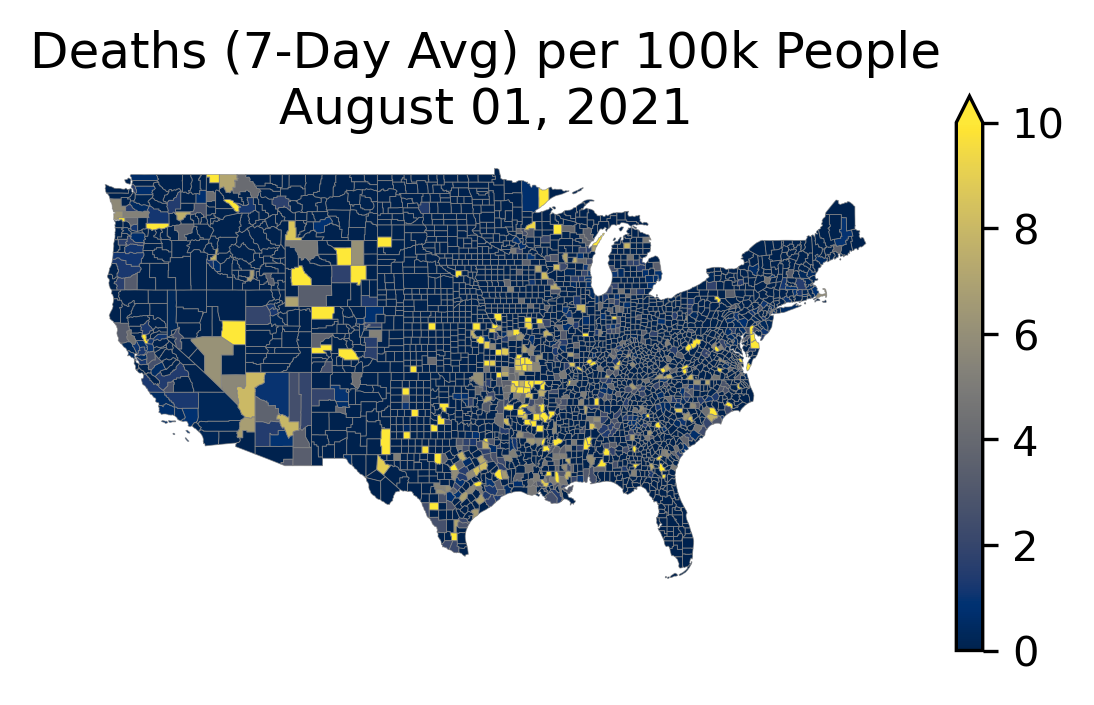}
	\caption{From left to right: contiguous United States county-level vaccination percentage, cases per 100,000 people, and deaths per 100,000 people, all on August 1st, 2021. Note that the counties in white in the ``Vaccination Percent'' figure on the left (several California counties and the entire state of Texas) did not report vaccination data to the Centers for Disease Control, and were therefore excluded in that figure.}
	\label{fig:covid_maps}
\end{figure*}

\section{Polar Parallel Coordinates Plots with Covid-19 Data in the United States}
\label{sec:covid}

Applying the techniques from Section \ref{sec:hive_plots}, we visually explored the trivariate relationship between vaccination rates, Covid-19 cases, and Covid-19 deaths for counties in the contiguous United States from March, 2021 through August, 2021. Specifically, we focused on the relative outcomes of the most-vaccinated and least-vaccinated counties. Vaccination data came from the Centers for Disease Control \cite{cdc2021vaccinations}, and Covid-19 cases and deaths data came from Johns Hopkins University \cite{clarke2021measuring}.

First, we considered these three variables in their geographic context for a single day of data---August 1st, 2021---in Figure \ref{fig:covid_maps}. There are certainly geospatial patterns here \textit{within} each map, but these maps don't lend themselves well to considering the multivariate relationship between the variables.

\begin{figure*}[h!]
	\centering
	\includegraphics[width=7in]{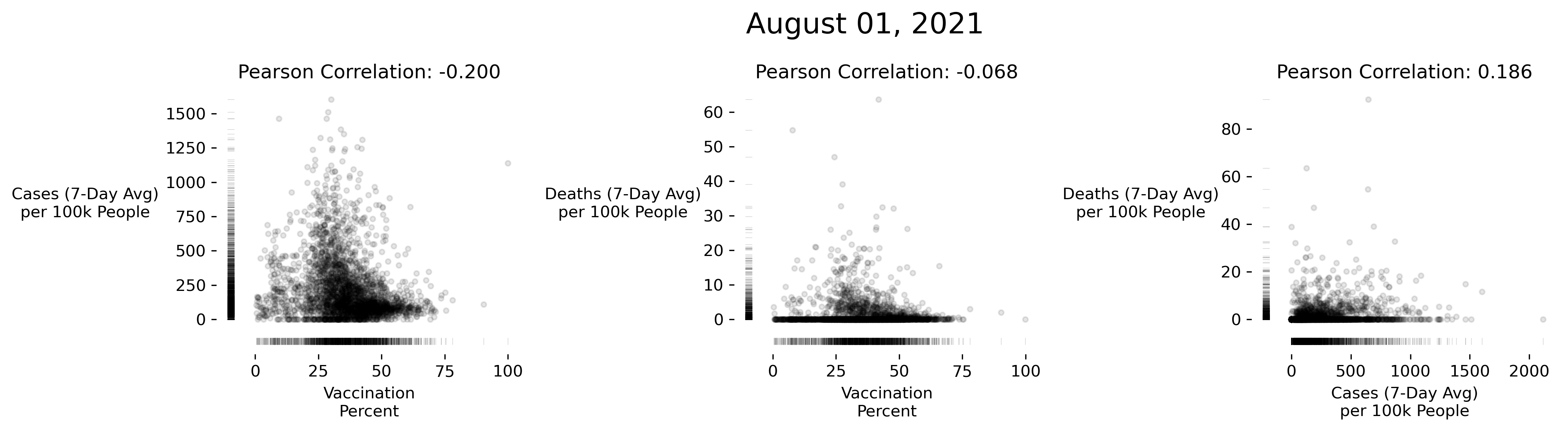}
	\qquad
	\includegraphics[width=4in]{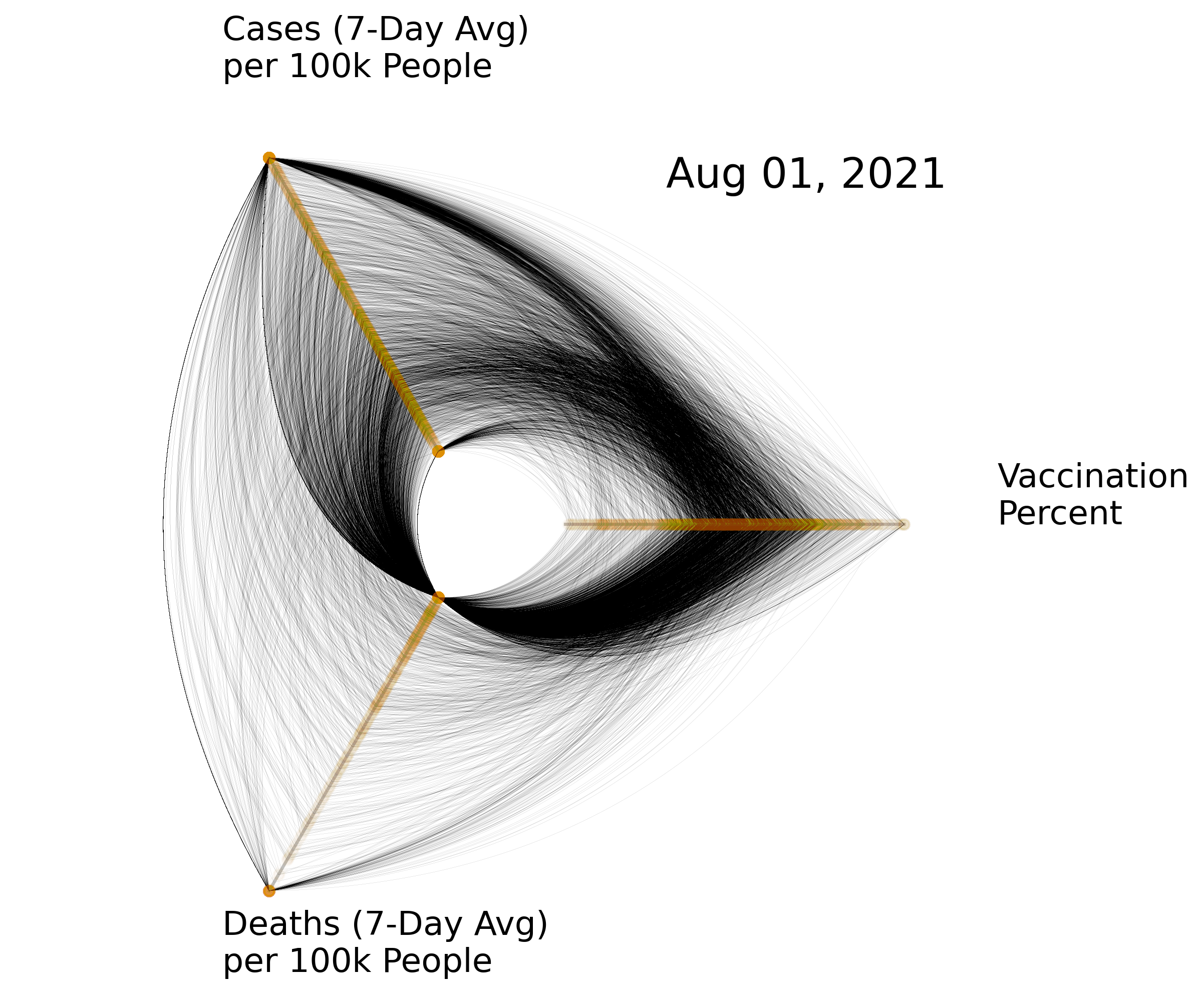}
	\caption{\textit{Top}: Dot-Dash Plots between the three possible bivariate combinations of contiguous United States county-level vaccination rates, cases per 100,000 people, and deaths per 100,000 people, all on August 1st, 2021 (the same data as shown in Figure \ref{fig:covid_maps}). The Pearson correlation between each pair of variables is reported above each plot. Univariate relationships can be viewed along the axes. At first glance, the authors found that these figures failed to suggest any particular hypotheses worth exploring. \textit{Bottom}: Polar Parallel Coordinates Plot of same data as above. Univariate density comparable to the axes of a Dot-Dash Plot is represented with the orange dots on each axis. Though not showing anything immediately obvious, this figure clearly shows \textit{heterogeneous relationships} between the variables that suggests further exploration. A natural way to drill-down on heterogeneity is to look at only the data with the highest and lowest values for one variable, for example exploring the behavior along different extremes of vaccination rates. Note that the minimum and maximum values on the axes of the Polar Parallel Coordinates Plot have been constrained to normalize outliers to a more tightly-bounded range.}
	\label{fig:covid_eda_no_color}
\end{figure*}

Next, we considered two different forms of comparison of the same data---bivariate Dot-Dash Plots and a P2CP---in Figure \ref{fig:covid_eda_no_color}. Perhaps the greatest surprise from the Scatterplots upon first glance is the lack of any apparent strong relationship between deaths and vaccination rates, and unfortunately, the Scatterplots offer little suggestion of the next step in a visual exploration of these data. The P2CP, on the other hand, shows an intriguing \textit{heterogeneous relationship} between deaths and vaccination rates reminiscent of the heterogeneous relationships considered in Figure \ref{fig:trivariate_corr}, where one could see visually distinctive behavior within a subset of the data despite minimal overall correlation.

\begin{figure*}[h!]
	\centering
	\includegraphics[width=7in]{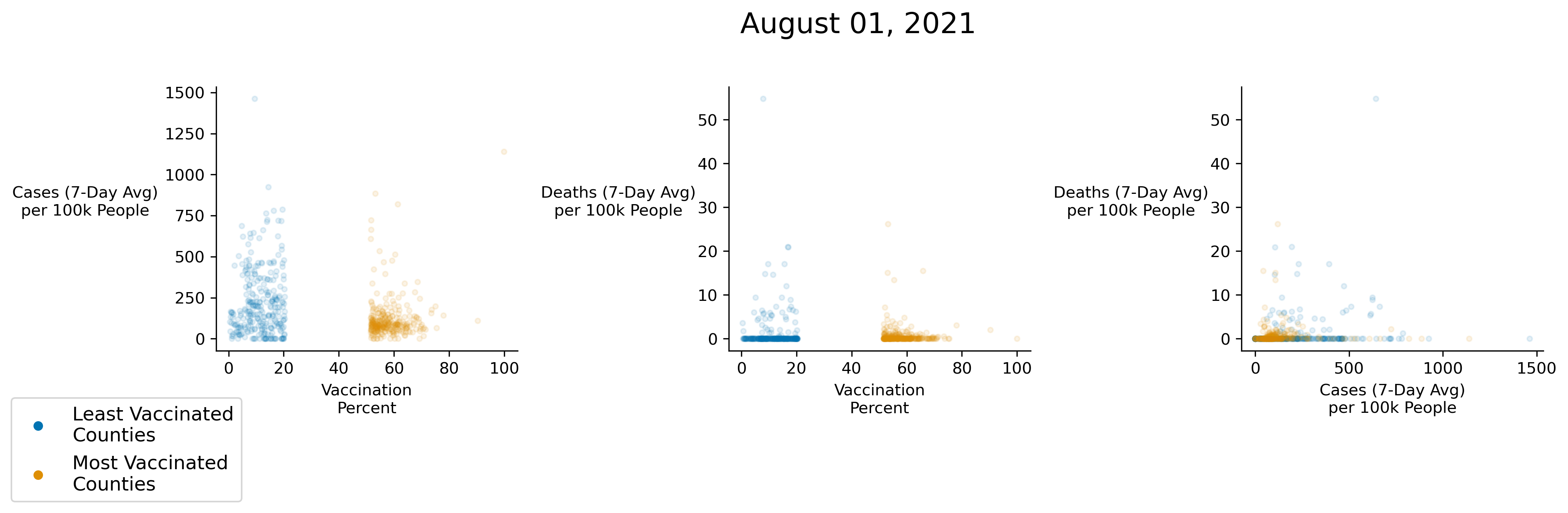}
	\qquad
	\includegraphics[width=4in]{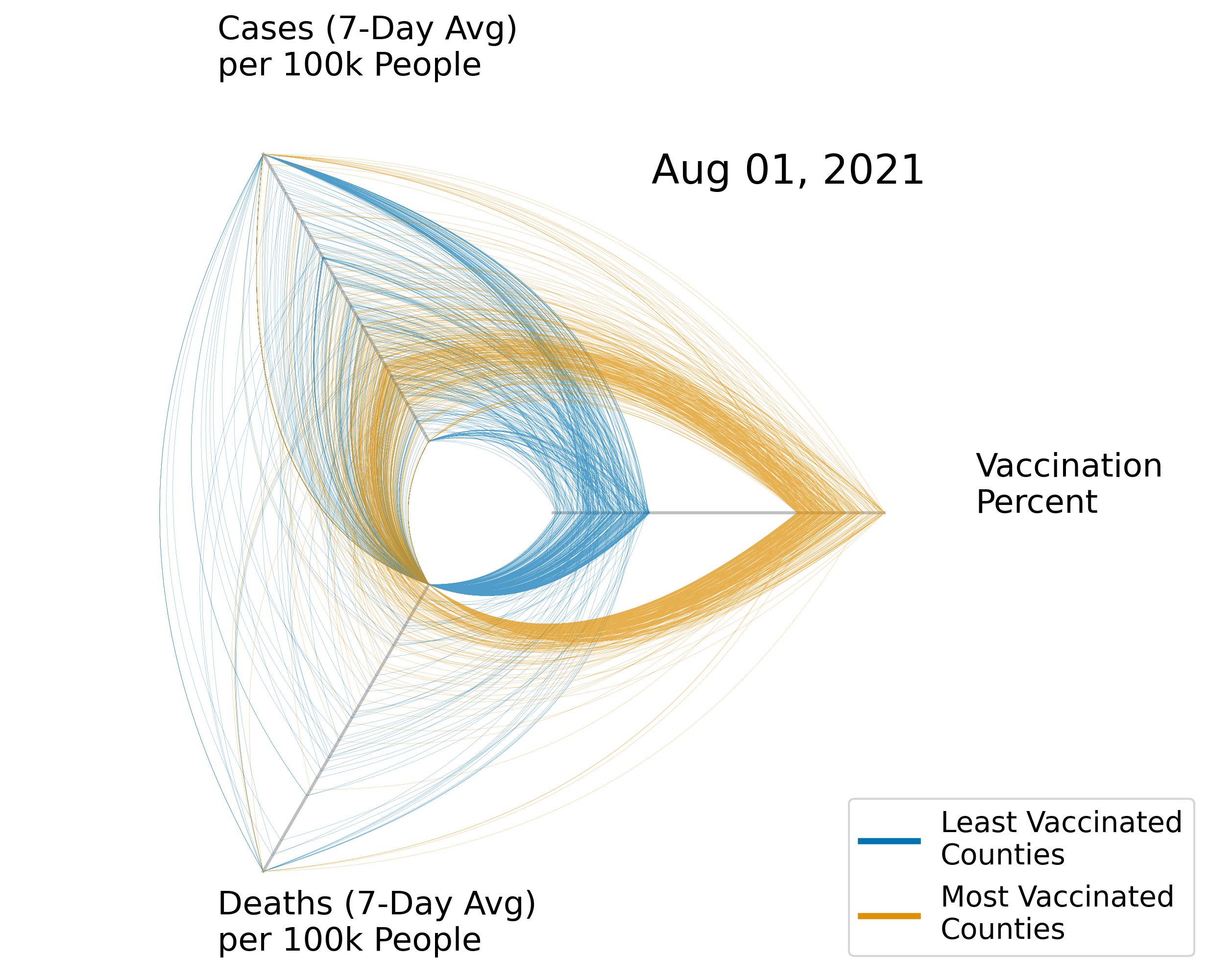}
	\caption{\textit{Top}: Scatterplots between the three possible bivariate combinations of contiguous United States county-level vaccination rates, cases per 100,000 people, and deaths per 100,000 people, all on August 1st, 2021. Data subset to the 10\% of counties with the highest vaccination percentage (orange) and the 10\% of counties with the lowest vaccination percentage (blue). \textit{Bottom}: Polar Parallel Coordinates Plot of the same subset of data as above.  Univariate density on the axes is removed in this figure.	Both the Scatterplots and the Polar Parallel Coordinates Plot allow for similar bivariate conclusions. Though both groups generally have low death rates, there are more outlier counties with high death rates among the least-vaccinated counties than the most-vaccinated counties. As for cases, though both groups have plenty of counties with low case rates, higher case rates are far more common among the least-vaccinated counties, whereas they are relative outliers among the most-vaccinated counties. Finally, cases and deaths seem to be positively correlated among some of the least-vaccinated counties, with a subset of those counties in addition to the vast majority of the most-vaccinated counties having little to no deaths regardless of case count. Note that the minimum and maximum values on the axes of the Polar Parallel Coordinates Plot have been constrained to normalize outliers to a more tightly-bound range.}
	\label{fig:covid_eda_color}
\end{figure*}

A natural starting point with heterogeneous behavior is to focus on just the extremes of a variable; to start, we will focus on the lowest and highest quantiles of vaccination percentage. In Figure \ref{fig:covid_eda_color}, we replicate Figure \ref{fig:covid_eda_no_color}, but we keep and color only the highest and lowest 10\% of counties by vaccination percentage. Several interesting patterns can be seen in this figure. First, there is a clear divergence in common behavior between the groups in terms of case rates, with less-vaccinated counties seeing more cases in general, though it should be noted that highly-vaccinated counties also have counties with high case rates. Second, both groups have relatively low death rates, but the less-vaccinated counties have noticeably more outlier counties with high death rates. Finally, we can see a fairly heterogeneous relationship between cases and deaths among the less-vaccinated group of counties. Namely, there appear to be two relationships---many of the less-vaccinated counties seem to observe a fairly standard positive correlation between cases and deaths, whereas another subset are following the trend of the more-vaccinated counties, that is, maintaining a low death rate regardless of case rate. This might be indicative of an omitted variable dividing outcomes in the less-vaccinated group of counties.\footnote{Variable behavior could be the result of anything from heterogeneous spread of more-infectious variants to demographic conditions (e.g. average age, obesity rates, etc. in counties). Once again, though, our P2CP visualization suggests hypotheses worthy of further exploration that are not visually suggested by comparable Scatterplots.}

\begin{figure*}[h!]
	\centering
	\includegraphics[width=7in]{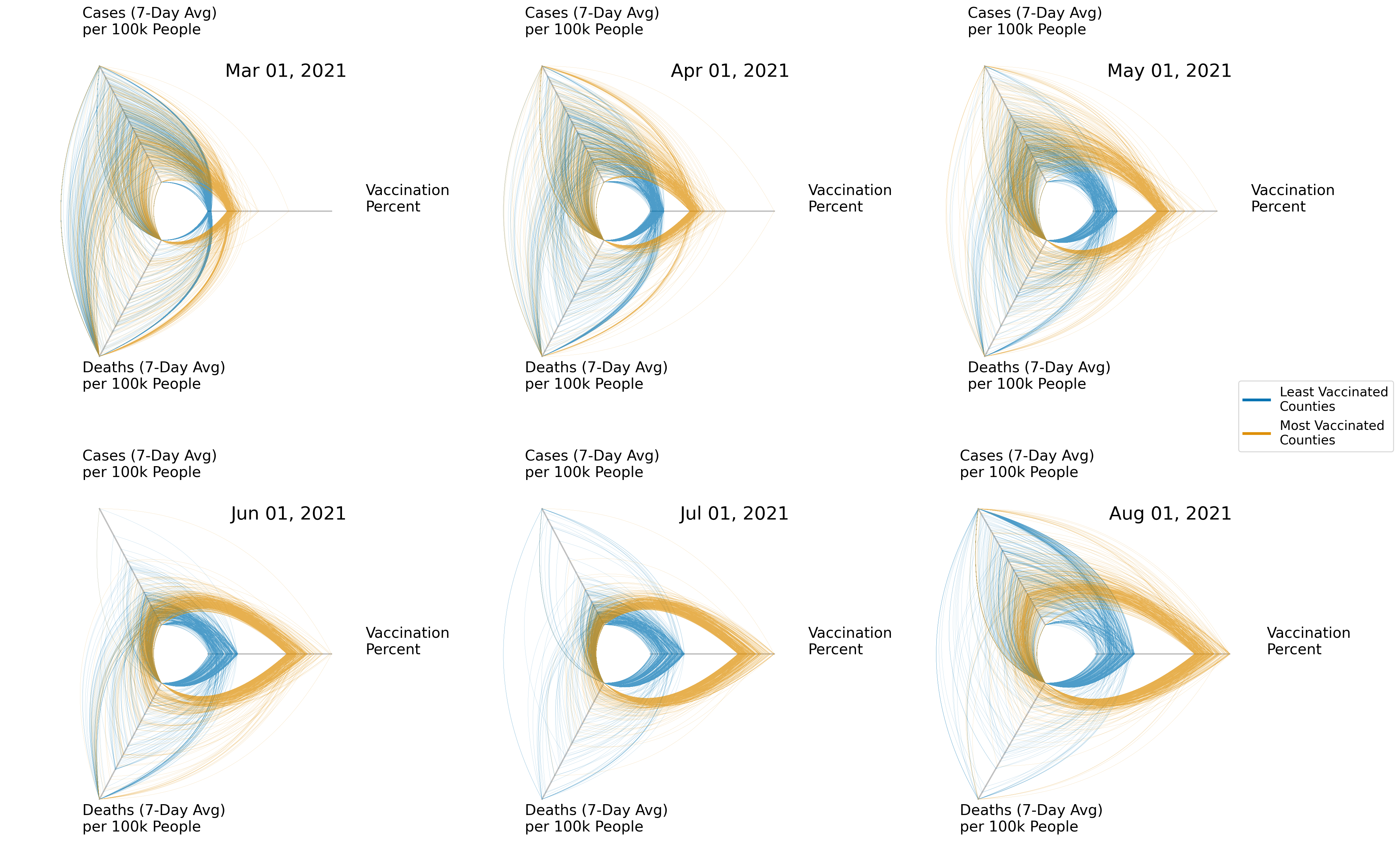}
	\caption{Polar Parallel Coordinates Plots between contiguous United States county-level vaccination rates, cases per 100,000 people, and deaths per 100,000 people, spanning six months from March 1st, 2021 through August 1st, 2021. Data subset to the 10\% of counties with the highest vaccination percentage (orange) and the 10\% of counties with the lowest vaccination percentage (blue). Univariate density on the axes is removed in each figure. The separations discussed with respect to Figure \ref{fig:covid_eda_color} become more pronounced over time. Note, the same Polar Parallel Coordinates Plot from Figure \ref{fig:covid_eda_color} is shown in the bottom right of this figure.}
	\label{fig:covid_hive_panel}
\end{figure*}

Taking advantage of the small multiples capability of P2CPs, we can look for \textit{trends} in these observed behaviors over time. In Figure \ref{fig:covid_hive_panel}, we find that the trends and separations discussed in Figure \ref{fig:covid_eda_color} are steadily converged on over time. Furthermore, we are able to explore these trends looking at only six plots as opposed to the eighteen that would be required were we instead to look at bivariate Scatterplots.

\section{Conclusion}
\label{sec:conclusion}

When augmented with techniques from HP network visualizations, P2CPs offer a compact, interpretable means for multidimensional visualization in flatland that lends itself well to plotting with small multiples. By taking advantage of the unique properties of 3-axis P2CPs in particular, we can collapse 3-dimensional data into two dimensions without sacrificing the visualization of any variable interactions. P2CPs offer a particularly good means for the exploration of multivariate, heterogeneous correlations, which makes this visualization technique a strong tool when working with socioeconomic data. Finally, P2CPs encourage hypothesis generation, making them an excellent starting point when beginning to analyze a dataset.

\section*{Acknowledgments}

This material is based upon work supported by the National Science Foundation under Grant No. 2029153. We are grateful to Tessa Johnson and Sam Voisin for technical discussion.

\clearpage

\bibliography{bibliography}

\begin{thebibliography}{10}

\bibitem{tufte2006beautiful}
E.~R. Tufte, {\em Beautiful evidence}, vol.~1.
\newblock Graphics Press Cheshire, CT, 2006.

\bibitem{chernoff1973use}
H.~Chernoff, ``The use of faces to represent points in k-dimensional space
  graphically,'' {\em Journal of the American statistical Association},
  vol.~68, no.~342, pp.~361--368, 1973.

\bibitem{d1885coordonnees}
M.~d'Ocagne, {\em Coordonn{\'e}es parall{\`e}les et axiales: M{\'e}thode de
  transformation g{\'e}om{\'e}trique et proc{\'e}d{\'e} nouveau de calcul
  graphique d{\'e}duits de la consid{\'e}ration des coordonn{\'e}es
  parall{\`e}les}.
\newblock Gauthier-Villars, 1885.

\bibitem{hewes1883scribner}
F.~W. Hewes, H.~Gannett, {\em et~al.}, ``Scribner's statistical atlas of the
  united states,'' 1883.

\bibitem{inselberg1985plane}
A.~Inselberg, ``The plane with parallel coordinates,'' {\em The visual
  computer}, vol.~1, no.~2, pp.~69--91, 1985.

\bibitem{heinrich2013state}
J.~Heinrich and D.~Weiskopf, ``State of the art of parallel coordinates.,'' in
  {\em Eurographics (State of the Art Reports)}, pp.~95--116, 2013.

\bibitem{von1877gesetzmassigkeit}
G.~Von~Mayr, {\em Die gesetzm{\"a}ssigkeit im gesellschaftsleben}, vol.~23.
\newblock De Gruyter Oldenbourg, 1877.

\bibitem{siirtola2009visual}
H.~Siirtola, T.~Laivo, T.~Heimonen, and K.-J. R{\"a}ih{\"a}, ``Visual
  perception of parallel coordinate visualizations,'' in {\em 2009 13th
  International Conference Information Visualisation}, pp.~3--9, IEEE, 2009.

\bibitem{lanzenberger2005exploring}
M.~Lanzenberger, S.~Miksch, and M.~Pohl, ``Exploring highly structured data: a
  comparative study of stardinates and parallel coordinates,'' in {\em Ninth
  International Conference on Information Visualisation (IV'05)}, pp.~312--320,
  IEEE, 2005.

\bibitem{hoffman2000table}
P.~E. Hoffman, {\em Table visualizations: a formal model and its applications}.
\newblock University of Massachusetts Lowell, 2000.

\bibitem{lanzenberger2003interactive}
M.~Lanzenberger, {\em The interactive stardinates: an information visualization
  technique applied in a multiple view system}.
\newblock PhD thesis, 2003.

\bibitem{radarchartcritique}
{From Data to Viz}, ``The radar chart and it's caveats.''
  \url{https://www.data-to-viz.com/caveat/spider.html}.
\newblock Accessed: 2021-08-03.

\bibitem{krzywinski2012hive}
M.~Krzywinski, I.~Birol, S.~J. Jones, and M.~A. Marra, ``Hive plots—rational
  approach to visualizing networks,'' {\em Briefings in bioinformatics},
  vol.~13, no.~5, pp.~627--644, 2012.

\bibitem{heinrich2011evaluation}
J.~Heinrich, Y.~Luo, A.~E. Kirkpatrick, H.~Zhang, and D.~Weiskopf, ``Evaluation
  of a bundling technique for parallel coordinates,'' {\em arXiv preprint
  arXiv:1109.6073}, 2011.

\bibitem{zhang2012network}
Z.~Zhang, K.~T. McDonnell, and K.~Mueller, ``A network-based interface for the
  exploration of high-dimensional data spaces,'' in {\em 2012 IEEE Pacific
  Visualization Symposium}, pp.~17--24, IEEE, 2012.

\bibitem{hurley2010pairwise}
C.~B. Hurley and R.~Oldford, ``Pairwise display of high-dimensional information
  via eulerian tours and hamiltonian decompositions,'' {\em Journal of
  Computational and Graphical Statistics}, vol.~19, no.~4, pp.~861--886, 2010.

\bibitem{anderson1936species}
E.~Anderson, ``The species problem in iris,'' {\em Annals of the Missouri
  Botanical Garden}, vol.~23, no.~3, pp.~457--509, 1936.

\bibitem{fisher1936use}
R.~A. Fisher, ``The use of multiple measurements in taxonomic problems,'' {\em
  Annals of eugenics}, vol.~7, no.~2, pp.~179--188, 1936.

\bibitem{dietzsch2009spray}
J.~Dietzsch, J.~Heinrich, K.~Nieselt, and D.~Bartz, ``Spray: A visual analytics
  approach for gene expression data,'' in {\em 2009 IEEE Symposium on Visual
  Analytics Science and Technology}, pp.~179--186, IEEE, 2009.

\bibitem{johansson2007depth}
J.~Johansson, P.~Ljung, and M.~Cooper, ``Depth cues and density in temporal
  parallel coordinates.,'' in {\em EuroVis}, vol.~7, pp.~35--42, Citeseer,
  2007.

\bibitem{barlow2004animator}
N.~Barlow and L.~J. Stuart, ``Animator: A tool for the animation of parallel
  coordinates,'' in {\em Proceedings. Eighth International Conference on
  Information Visualisation, 2004. IV 2004.}, pp.~725--730, IEEE, 2004.

\bibitem{theron2006visual}
R.~Theron, ``Visual analytics of paleoceanographic conditions.,'' in {\em IEEE
  VAST}, pp.~19--26, Citeseer, 2006.

\bibitem{countyUnemploymentData}
{United States Bureau of Labor Statistics}, ``Labor force data by county,
  annual averages.'' \url{https://www.bls.gov/lau/#cntyaa}.
\newblock Accessed: 2021-08-25.

\bibitem{countyPovertyData}
{United States Census Bureau}, ``Saipe state and county estimates.''
  \url{https://www.census.gov/programs-surveys/saipe.html}.
\newblock Accessed: 2021-08-25.

\bibitem{li2010judging}
J.~Li, J.-B. Martens, and J.~J. Van~Wijk, ``Judging correlation from
  scatterplots and parallel coordinate plots,'' {\em Information
  Visualization}, vol.~9, no.~1, pp.~13--30, 2010.

\bibitem{johansson2008perceiving}
J.~Johansson, C.~Forsell, M.~Lind, and M.~Cooper, ``Perceiving patterns in
  parallel coordinates: determining thresholds for identification of
  relationships,'' {\em Information Visualization}, vol.~7, no.~2,
  pp.~152--162, 2008.

\bibitem{kuang2012tracing}
X.~Kuang, H.~Zhang, S.~Zhao, and M.~J. McGuffin, ``Tracing tuples across
  dimensions: A comparison of scatterplots and parallel coordinate plots,'' in
  {\em Computer Graphics Forum}, vol.~31, pp.~1365--1374, Wiley Online Library,
  2012.

\bibitem{tufte2001visual}
E.~Tufte, ``The visual display of quantitative information,'' 2001.

\bibitem{cdc2021vaccinations}
{Centers for Disease Control}, ``Covid-19 vaccinations in the united states,
  county.'' \url{https://data.cdc.gov}.
\newblock Accessed: 2021-08-08.

\bibitem{clarke2021measuring}
J.~M. Clarke, A.~Majeed, and T.~Beaney, ``Measuring the impact of covid-19,''
  2021.

\end{thebibliography}
\bibliographystyle{ieeetr}

\onecolumn

\section*{Appendix}

\begin{figure*}[h!]
	\centering
	\includegraphics[width=7in]{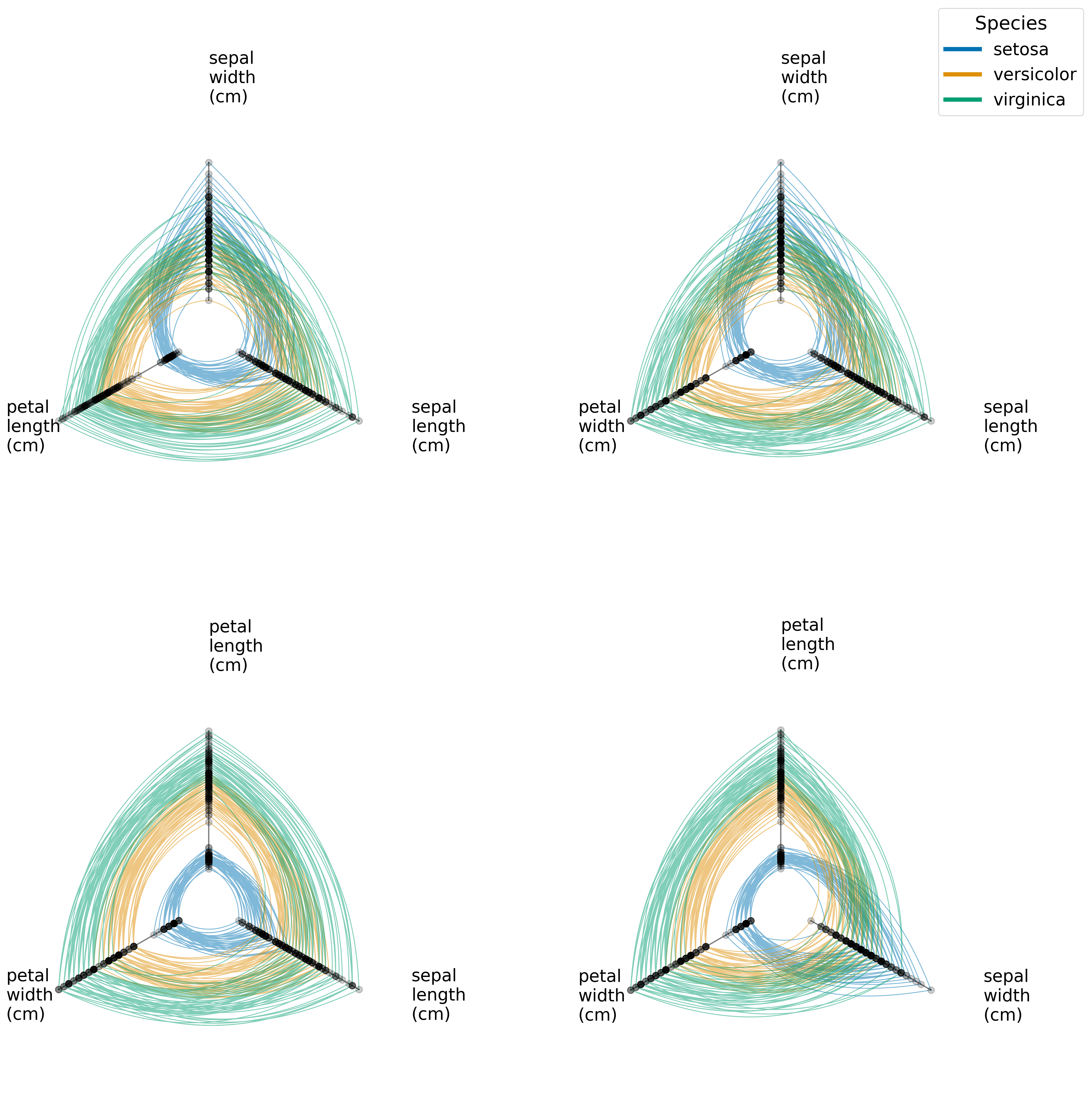}
	\caption{Four instances of three-axis Polar Parallel Coordinates Plots showing all possible combinations of three variables in the Iris dataset. Note that this figure has comparable information content to Figure \ref{fig:iris_scatter}, but only requires 4 plots instead of 10.}
	\label{fig:iris_p2cp_panel}
\end{figure*}

\end{document}